\documentstyle[12pt,epsf]{article}

\begin{document}

\begin{titlepage}

\begin{flushright}
{\normalsize UTHEP-348\\
UTCCP-P-19\\
February, 1997
}
\end{flushright}

\vspace*{0.5cm}
\begin{centering}
{\Large \bf 
Scaling in SU(3) Pure Gauge Theory\\
\vspace{2mm}
with a Renormalization Group\\
\vspace{2mm}
Improved Action
}
\vspace{0.8cm}

{\large
Y.\ Iwasaki\rlap,$^{\rm a,b}$ K.\ Kanaya\rlap,$^{\rm a,b}$
T.\ Kaneko\rlap,$^{\rm a}$
and T.\ Yoshi\'e$^{\rm a,b}$
}
\vspace{0.5cm}

{\it
$^{\rm a}$
Institute of Physics, University of Tsukuba,
Ibaraki 305, Japan\\
$^{\rm b}$
Center for Computational Physics, University of Tsukuba,\\
Ibaraki 305, Japan\\
}

\end{centering}

\vspace{1cm}\noindent
{
We study the scaling properties of
the static quark potential and
the ratio of the critical temperature $T_c$
to the square root of the string tension $\sigma$
in the SU(3) pure gauge theory
using a renormalization group improved action.
We first determine the critical coupling $\beta_c$
on lattices with temporal extension $N_t=3$, 4, and 6,
and then calculate the static quark potential
at the critical couplings
on lattices at zero temperature.
We note that the static quark
potentials obtained are rotationally invariant with errors of
at most 1 -- 2 \% in all the three cases, and
that the potential $V(R)$ in physical units
scales in the whole region of $R$ investigated.
The values of $T_{c}/\sqrt{\sigma}$ for the three cases
in the infinite volume limit
are identical within errors.
We estimate the value in the continuum limit to be
$T_{c}/\sqrt{\sigma} = 0.656(4)$,
which is slightly larger than
the value in the continuum limit
from the one-plaquette action, 0.629(3).
}

\vfill \noindent

\end{titlepage}

\section{Introduction}

In numerical studies of lattice QCD,
it is important to control and reduce
finite lattice spacing effects.
Several improved actions have been proposed
for this purpose and some of them have been tested
for the scaling behavior of the critical temperature $T_c$ 
of the finite temperature deconfining transition 
\cite{SA,FA,SLW,BSATcs,UkawaLat96}.

In this work we study the scaling properties of 
the static quark potential and 
the ratio of the critical temperature to 
the square root of the string tension $\sigma$, $T_{c}/\sqrt{\sigma}$,
in the SU(3) pure gauge theory,  
using a renormalization group (RG) improved action \cite{RGIA}:
\begin{equation}
S_g^{\rm IM} = {\beta \over 6}\, \{c_0 \sum (1\times 1 {\rm \ loop})
               + c_1 \sum (1\times 2 {\rm \ loop})\}
\label{eq:action} 
\end{equation}
with $c_1=-0.331$ and $c_0=1-8c_1$, where $\beta = 6/g^2$ 
($g$ is the gauge coupling). 
In Eq.(\ref{eq:action}), 
the loops are defined by the trace of ordered product of
link variables and each oriented loop appears once in the sum.

This paper is organized as follows.
First we determine the critical coupling $\beta_c$'s
for the finite temperature deconfining phase transition
on $9^{3}\times3$, $12^3\times 4$ and $18^3\times 6$ lattices
in Sec.~2.
We also perform simulations 
on $12^3\times3$, $15^3\times3$, $16^3\times4$,
and $20^3\times4$ lattices
for a finite size scaling study. 
Then the quark potentials at the three $\beta_c$'s
are calculated from smeared Wilson loops
on $9^3 \times 18$,
$12^3\times 24$ and $18^3\times 36$ lattices, respectively, in Sec.~3.
The string tension is extracted from the quark potential
assuming that the potential takes 
a form of a sum of a Coulomb term and a linearly rising 
potential.
In Sec.~4, scaling behavior of the quark potential and
that of the ratio $T_{c}/\sqrt{\sigma}$ are examined.
Finally, the value of the ratio $T_{c}/\sqrt{\sigma}$
in the continuum limit and in the infinite volume limit is 
estimated.

\section{Critical coupling $\beta_c$}

In order to determine the critical coupling $\beta_c$ 
for the finite temperature
phase transition, 
we perform simulations 
on $9^{3}\times3$, $12^{3}\times4$ and $18^{3}\times6$ lattices. 
The critical temperature $T_c$ is given by $T_c=1/(a N_t)$,
where $N_t$ is the linear extension of the lattice in the temporal
direction 
and $a$ is the lattice spacing at the critical coupling.
Note that 
the physical spatial volumes are identical for all the three cases,
$(N_s a)^3= (3 N_t a )^3 = (3 / T_c)^3$,
where $N_s$ is the linear extension of the lattice in the spatial 
direction.

We also perform simulations 
on lattices with different spatial volumes 
for an estimation of the infinite volume limit of $\beta_c$
using finite size scaling analyses.
The previous results for the case of the standard one-plaquette
action on spatially large lattices \cite{fukugita,StdNt46} indicate 
that extrapolations from small lattices 
with the aspect ratio $N_s/N_t \leq 3$ result in 
sizable systematic errors in the values of $\beta_c$
in the infinite volume limit.
Therefore, we restrict ourselves to lattices $N_s/N_t \geq 3$
in this paper. We perform simulations 
on $12^3\times3$, $15^3\times3$, $16^3\times4$, and $20^3\times4$ 
lattices for finite size analyses.
We reserve finite size study of $N_t=6$ lattices for future 
investigation.

Gauge fields are updated 
by a Cabibbo-Marinari-Okawa pseudo heat bath algorithm with 8 hits 
both for the simulations at finite temperatures and at zero 
temperature discussed in the next section. 
The simulation parameters are compiled 
in Table~\ref{tab:FTPTpara}.
We measure Wilson loops and Polyakov line every 10 sweeps.
Their expectation values are summarized 
in Tables~\ref{tab:FTPTmeas1} - \ref{tab:FTPTmeas5}.
(For the deconfinement fraction, see below.)

The values of the critical coupling $\beta_c$
are determined as the peak location
of the susceptibility $\chi$
of the Z(3) rotated Polyakov line $\Omega$:
\begin{equation}
  \chi = \langle \;  \Omega ^2 \; \rangle
        -\langle \;  \Omega    \; \rangle^2
\end{equation}
\begin{equation}
  \Omega = \left\{
  \begin{array}{lll}

    \mbox{Re} \;
    P\exp\left[-\frac{2}{3}\pi\right]  \;
    \; &
    ;\;\;\;
      \mbox{arg}P
      \in \left[\pi/3,\pi\right)
    \\
    \mbox{Re}  \; P \;  \; &
    ;\;\;\;
      \mbox{arg}P
      \in \left[-\pi/3,\pi/3\right)
    \\
    \mbox{Re}  \;
    P\exp\left[\frac{2}{3}\pi\right] \;
     \;&
    ;\;\;\;
      \mbox{arg}P
      \in \left[-\pi,-\pi/3\right) {_,}
  \end{array}
\right.
\end{equation}
where $P$ is  
the spatially averaged timelike Polyakov line
\begin{equation}
  P = {\frac{1}{N_s^3}}\sum_{\bf{x}}
  \frac{1}{3}\makebox{Tr}\left[
  \prod_{t=1}^{N_t}U{_4}\left({\bf{x}},t\right)
  \right]_.
\end{equation}

The results of the susceptibility 
calculated using the spectral density method \cite{SDM}
on the $9^3\times3$, $12^3\times4$, and $18^3\times6$ lattices 
are shown in Fig.~\ref{fig:suscep}.
The results obtained
at several simulation points
are consistent with each other within the errors
and form a clear peak structure.
The value of $\beta_c$ is determined from 
the data at the $\beta$ which is the closest to $\beta_c$.
The errors are estimated using a single-elimination jack knife method. 
The bin size in the jack knife method is determined by investigating
the bin size dependence of the errors of
$\Omega $, shown in Fig.~\ref{fig:pjne}.
We note that the jack-knife errors of $\beta_c$'s are stable 
for the bin size larger than those adopted,
as shown in Fig.~\ref{fig:bcjne}.
The values of $\beta_c$'s and their jack-knife errors
are summarized in Table~\ref{tab:betac}.

There are several alternative definitions of $\beta_c$ on finite 
lattices. 
A popular method is to 
measure the ``deconfinement fraction'' $r$ given by
$r = (3p - 1)/2$
where $p$ is the probability
such that $\arg P \in 
(-\pi/9,\pi/9)$, $(2\pi/3-\pi/9,2\pi/3+\pi/9)$, 
or $(-2\pi/3-\pi/9,-2\pi/3+\pi/9)$, 
and to define $\beta_c$ as a point where $r$ takes a given value.
Our results of $r$ as a function of $\beta$ for the case of 
the aspect ratio $N_s/N_t=3$ are shown in Fig.~\ref{fig:dcf}.
See also Tables~\ref{tab:FTPTmeas1}-\ref{tab:betac}.
We find that the deconfinement fraction $r$ is approximately 0.75 
at $\beta_c$ determined from the peak location of the 
susceptibility,
as summarized in Table~\ref{tab:betac}.
We note that this fact for the deconfinement fraction 
is also realized in 
data~\cite{StdNt46} 
obtained for the standard one-plaquette action on large lattices 
with high statistics (see Table~\ref{tab:stdr}). 
The condition $r = 3/4$ is the
criterion taken in Ref.~\cite{gottlieb}
for the determination of $\beta_c$.
(See also the discussions in Refs.~\cite{christ86,toussaint}.)
However, the volume dependence of the corrections of $\beta_c$ 
to the infinite volume limit is not known in this case.%
\footnote{
The value $r=3/4$ corresponds to the case that
the four peaks of the histogram
of $P$ in the complex plane have the same volume 
fraction~\cite{toussaint},
assuming uniformity of the distribution in terms of $\arg P$ in the
confining phase. 
For the $q$-state Potts models with large $q$, the value of $\beta_c$
which corresponds to the case where $(q+1)$ peaks have the same 
volume fraction
is shown to yield the correct infinite volume value of $\beta_c$
up to exponentially suppressed corrections \cite{Borgs91}.
However, in the SU(3) gauge theory, uniformity of distribution 
in terms of $\arg P$ in the confining phase is not well satisfied. 
Therefore, $r=3/4$ does not 
strictly correspond to the case of equal weight of four peaks.
Thus, in contrast to the case of $\beta_c$ from the peak location 
of the susceptibility, 
no rigorous scaling relation is known for the
$\beta_c$ determined from the deconfinement fraction.
In practice, when we adopt $\beta_c$ determined from $r=3/4$ and 
assume either a linear volume dependence or an exponential volume 
dependence, 
we obtain a result for  $T_c/\sqrt{\sigma}$ in the continuum limit
which agrees, within errors, with that derived in the text 
using $\beta_c$ from the peak location of the susceptibility.
}
On the other hand, a scaling relation is well established for the
$\beta_c$ determined from the peak location of the
susceptibility.
Therefore, 
we concentrate on $\beta_c$ determined from the peak location of 
the susceptibility for finite size scaling analyses. 

In the following, 
we denote the $\beta_c$ on the $9^3\times3$, 
$12^3\times4$, and $18^3\times6$ lattices
as $\beta_c(9^3\times3)$, 
$\beta_c(12^3\times4)$, and $\beta_c(18^3\times6)$,
respectively.

\section{String tension}

We evaluate the string tensions
at $\beta=\beta_c(9^3\times3)$, $\beta_c(12^3\times4)$, 
and  $\beta_c(18^3\times6)$ on lattices at zero temperature:
$9^3\times18$, $12^3\times24$, and $18^3\times36$ lattices, 
respectively.
Note that the spatial sizes of the lattice
are the same as those for the finite temperature simulations
in all the three cases. The ratio $N_t/N_s$ is also fixed to 2.
The simulation parameters are summarized in Table~\ref{tab:STpara}.
After thermalization sweeps,
we measure Wilson loops every 200 sweeps.
The spatial paths of the loops are formed
by connecting one of the spatial vectors
shown in Fig.~\ref{fig:bus}.

In order to extract the ground state contribution 
to the potential,
we adopt the smearing technique proposed
in Ref.\cite{StdST}:
Each spatial link $U$ is replaced with 
an SU(3) matrix $U_{\rm new}$
which maximizes
$\mbox{ReTr}\left[ F U_{\rm new} \right]$,
with $F$ being the sum of the spatial staple products of link variables 
around $U$.
We perform this procedure up to
10, 30 and 40 steps on the $9^3\times18$, $12^3\times24$
and $18^3\times36$ lattices, respectively.
Measurements are carried out every smearing step 
on the $9^3\times18$ and 
every 2 smearing steps on the other lattices.
With this smoothing procedure
the behavior of the effective mass 
\begin{equation}
m_{\rm eff} = \mbox{log}\left[ W(R,T) / W(R,T+1) \right]
\end{equation}
in terms of $T$ 
is much improved, especially for large $R$ 
as shown in Fig.~\ref{fig:em}.

In the following, we discuss separately the results of
the potential $V(R)$ at $\beta_c$ for $N_t=4$ and 6, and that for 
$N_t=3$, because in the former case we are able to extract 
the coefficient of the Coulomb term 
by a straightforward fitting procedure with examining 
the stability of the fit,
while in the latter case 
it is hard to fix it solely from the data
due to a small number of the data points 
caused by the coarseness of the lattice at $\beta_c(9^3\times3)$ 
(see discussions below).

\subsection{Results at $\beta_c(12{^3}\times4)$
            and $\beta_c(18{^3}\times6)$}

The potential $V(R)$ and the overlap function $C(R)$ 
are extracted 
by a fully correlated fit of Wilson loops to the form
\begin{equation}
  W(R,T) = C(R) {\cdot} \mbox{ exp}\left[ -V(R) {\cdot} T \right].
  \label{eqn:fit1}
\end{equation}

The fitting range is determined by examining carefully $\chi{^2}/df$ 
and stability of $V(R)$ against the smearing step.
Fig.~\ref{fig:cVstb} shows the results of $\chi{^2}/df$ 
and $V(R)$ versus the smearing step at $R=4.0$
for the case of $\beta_c(12{^3}\times4)$.
When we take the fitting range $T=3-5$, 
we find that $\chi{^2}/df$
\raise0.3ex\hbox{$<$\kern-0.75em\raise-1.1ex\hbox{$\sim$}} 1 
and $V(R)$ is quite stable
after four smearing steps, 
while the choice of the fitting range $T=2-4$ 
leads to $\chi{^2}/df$ much larger than 1 and
a significant variation of $V(R)$ against the smearing step.
We find 
that the choice of the fitting range $T=3-5$ leads to
reasonable $\chi^2/df$ and stability of $V(R)$ against the
smearing step
for all $R$ except $2\sqrt{6}$
(where $\chi^2/df$ takes a little large value $\sim 2.5$,
though the stability is satisfied).
This stability implies that 
the contamination from excited states is negligible small.
Therefore, we take the fitting range $T=3-5$ for the 
data at $\beta_c(12{^3}\times4)$.
The $T$ range $4-7$ at $\beta_c(18{^3}\times6)$
is determined in a similar way.

We determine the optimum number of smearing steps 
for each $R$ in such a way that $C(R)$ takes the largest value
under the condition $C(R) \le 1$ which we call 
the ``optimum smearing step''.
We note that $\chi^2/df$ is stable 
(\raise0.3ex\hbox{$<$\kern-0.75em\raise-1.1ex\hbox{$\sim$}} 1)
against a variation of the smearing step when 
$C(R) \simeq 1$.
The optimum smearing steps thus determined are 
about 8 at $\beta_c(12{^3}\times4)$,
and are distributed from $12$ to $40$ at $\beta_c(18{^3}\times6)$
(see Tables \ref{tab:STmeas1} and \ref{tab:STmeas2}).%
\footnote{
We find that the value of $C(R)$ for $R=1.0$ -- 2.0 on 
the $18^3\times36$ lattice is greater than 1 at all
smearing steps $\le 40$.
We have checked using 20 configurations that more than $60$ 
smearing steps 
are needed to get $C(R) \le 1$ for these $R$'s.
Because we do not use these small loops for the fit of the potential,
we stop the smearing steps at 40 times.}
We take the value of $V(R)$ at the optimum smearing step.
The systematic error due to the choice of the smearing step
is much smaller than the statistical error,
because the value of $V(R)$ is stable against the smearing step
as mentioned above, and therefore we neglect it in the following.

The values for $V(R)$
are summarized in Tables \ref{tab:STmeas1} and \ref{tab:STmeas2}.
Statistical errors
are estimated by the jack-knife method 
with bin size 1.
Note that measurements are performed every 200 sweeps.
We confirm that the errors are quite stable against the bin size.

The string tension is determined 
by fitting $V(R)$ to the rotationally invariant ansatz
\begin{equation}
  V(R) = V_0 - {\alpha \over R} + {\sigma_{\rm lat}}R,
  \label{eqn:fit2}
\end{equation}
where $\sigma_{\rm lat} = \sigma a^2$ 
is the string tension in lattice units. 
We take into account the correlations among $V(R)$ at different $R$ 
using the error matrix derived from 
those for $W(R,T)$.
The fitting ranges we take are 
\begin{equation}
  \begin{array}{ll}
    R = \sqrt{6}\;-\;4\sqrt{2}  & (12{^3}{\times}24), \\
    R = 2\sqrt{3}\;-\;4\sqrt{5} & (18{^3}{\times}36). \\
  \end{array}
\end{equation}
These ranges ($R_{\rm min}$ -- $R_{\rm max}$) are
determined by investigating the stability of fits 
and the value of $\chi^2/df$ as explained in the following.
As we increase $R_{\rm min}$, 
instability of the fit first appears in the result 
of $\alpha$, while the results of $V_0$ and $\sigma_{\rm lat}$ 
are stable.
The error of $\alpha$ becomes abruptly large as $R_{\rm min}$ 
increases: e.g. at $\beta_c(12{^3}\times4)$
with $R_{\rm max}=4\sqrt{2}$ fixed,
$\alpha= 0.332(11),0.295(14),0.154(101)$, and -0.040(121) for 
$R_{\rm min} = \sqrt{5}$, $\sqrt{6}$, $2\sqrt{2}$, and 3.0, 
respectively.
Therefore, we restrict candidates for $R_{\rm min}$ 
to those for which the error of $\alpha$ is less than 50\% 
of the central value. 
We find that $\chi^2/df$ is stable and $\sim 1$ for 
$5 \le R_{\max} < 6$ [$7 \le R_{\max} < 9$ ] at
$\beta_c(12{^3}\times4)$ [$\beta_c(18{^3}\times6)$]  
which we take as the candidates for $R_{\rm max}$. 
The fitting range is determined by the condition that
$\chi^2/df$ takes a value nearest to 1 
in all the combinations of the candidates 
for $R_{\rm min}$ and $R_{\rm max}$.
The values of $\chi^2/df$ are 1.5 and 1.2
at $\beta_c(12{^3}\times4)$ and $\beta_c(18{^3}\times6)$, respectively,
for the $R_{\rm min}$ and $R_{\rm max}$ adopted.
We have checked that the results of $\alpha$ and $\sigma_{\rm lat}$ 
are stable for all candidates of $(R_{\rm min},R_{\rm max})$
which satisfy
\begin{equation}
  \begin{array}{ll}
2.0 \le R_{\rm min} \le \sqrt{6},\;\;\;\;\; 5.0 
\le R_{\rm max} \le 6.0 &
(12^3\times24) \\
3.0 \le R_{\rm min} \le 3\sqrt{2},\;\;\;7.0 
\le R_{\rm max} \le 9.0 &
(18^3\times36). \\
  \end{array}
\end{equation}

Note that the changes of the fitting ranges of $R$
at these two $\beta$'s
are consistent with the change of the scale between 
$\beta=\beta_c(12{^3}{\times}4)$ and $\beta_c(18{^3}{\times}6)$,
that is, the ratio of 4 to 6.

The results of $V_0$, $\alpha$, $\sigma_{\rm lat}$,
and their jack knife errors are summarized 
in Table~\ref{tab:FitRst}.
The values of $V(R)$ are plotted in Fig.~\ref{fig:VvsRds},
where different symbols
correspond to different units of spatial path of Wilson loops. 
The values of $V(R)$ obtained
from six types of Wilson loops 
are excellently fitted to 
the rotationally invariant form, 
Eq.(\ref{eqn:fit2}).
The deviations of the data at $\beta_c({12^3}\times4)$
from the fitted curve are less than 2\% and the average of them is 
about 0.4\%.
For the data at $\beta_c({18^3}\times6)$,
the deviations are at most 1\% with the average about 0.3\%.

We note that the results of $\alpha$ are consistent with a constant 
within the errors. 
The resulting $\alpha \simeq 0.296$ is slightly larger 
than $\pi/12 \simeq 0.262$ derived in a string model \cite{pi/12}.
We also perform fits with the value of $\alpha$
fixed to $\pi/12$. Then the values obtained are 
$\sigma_{\rm lat} = 0.1527(14)$ and $0.0667(6)$ 
at $\beta_c(12{^3}\times4)$ and $\beta_c(18{^3}\times6)$, 
respectively.
The values for the ratio $T_c/\sqrt\sigma$ using these results are
consistent with our final results using the values in 
Table~\ref{tab:FitRst}
within one standard deviation.

\subsection{Results at $\beta_c(9{^3}\times3)$}

We obtain the potential $V(R)$ at $\beta_c(9^3\times3)$  by fitting
$W(R,T)$ to the form Eq.(\ref{eqn:fit1}) 
with the fitting range $T=2\;-4\;$.
The fits with this fitting range have desirable properties
similar to those at the other two $\beta_c$'s discussed in the 
preceding subsection;
reasonable $\chi{^2}/df$ and
stability of $V(R)$ against the smearing step.

When we make a fit of the potential to the form Eq.(\ref{eqn:fit2}),
we find that the $R_{\rm min}$ dependence of $\alpha$ is stronger 
than the cases discussed in the previous subsection,
while the fits are quite stable against $R_{\max}$ like in 
the previous cases. 
This is due to the fact that we have only small number of data points 
at small $R$ caused by the coarseness of the lattices 
at $\beta_c(9^3\times3)$. 
Therefore a small deviation 
from the rotational invariance at $R=R_{\rm min}$ sometimes affects 
the value of $\alpha$ sizable. 
As a result, we are not able to find an $R_{\rm min}$ region 
for which $\alpha$ is stable.

Therefore, we perform two kinds of fits at $\beta=\beta_c(9^3\times3)$:
In the first fit, we fix the value of $\alpha$ to the average 
value 0.296 of those at the other two $\beta_c$'s 
which are constant within the errors.  
We set the fit range to be $R=2\;-\;2\sqrt{5}$
so that the physical $R$ range is consistent with 
the ranges at $\beta_c(12{^3}{\times}4)$
and $\beta_c(18{^3}{\times}6)$.
As shown in Fig.~\ref{fig:VvsRdsNt3}, the fit well reproduces the data 
even at $R<R_{\rm min}$.
In the other fit, we perform fit without fixing the value of $\alpha$, 
for the ranges $R_{\rm min}=\sqrt{2},\sqrt{3}$, and $2$ and 
$R_{\rm max} = 2\sqrt{5}$. 
These values of $R_{\rm min}$ in physical units correspond to 
those at the other two $\beta_c$'s 
for which the stability of $\alpha$ is observed.

We take the results of the former fit with $\alpha$ fixed as
the central values of $\sigma_{\rm lat}$ and $V_0$. 
The statistical errors are 
obtained by the jackknife method with bin size 1. 
We then take the upper bounds and lower bounds 
of $V_0$ and $\sigma_{\rm lat}$ 
obtained by the fits $\alpha$ unfixed,
as systematic errors.
The results of $V_0$ and $\sigma_{\rm lat}$ with the errors 
are given in Table~\ref{tab:FitRstNt3}.
The potential data are
shown in Fig.~\ref{fig:VvsRdsNt3}
together with its fit curve ($\alpha$ fixed to 0.296).
The deviations from the fit are at most 2\% and the average of them is
about 0.5\%, which indicates that
the rotational invariance is well restored 
even at this small value of $\beta$.

We also perform a fit with $\alpha$ fixed to $\pi/12$ to find 
$\sigma_{\rm lat} = 0.2607(16)$. 
The ratio $T_c/\sqrt\sigma$ using this result is
consistent with our final result using the value 
in Table~\ref{tab:FitRstNt3}
within the errors.

\section{Scaling properties}

In Fig.~\ref{fig:VvsRTc}, the values of $V_{\rm phys}/T_c$ 
are shown as a function of $R a \cdot T_c$, 
where $V_{\rm phys}=V/a$ is the potential in physical units. 
We note that the data on all the lattices
are in excellent agreement in the whole $R a \cdot T_c$ region.
This implies scaling of our potential data in the range of $\beta$
values investigated.
It might be emphasized again that the deviation of the data
from the rotationally invariant fit is at most 2 \% for the $N_t=3$
and 4 cases and 1 \% for the $N_t=6$ case.

Using the results presented in the preceding section,
we obtain the values of $T_c/\sqrt{\sigma}$ 
on the lattices with finite spatial volume $9^3$, $12^3$, and $18^3$, 
which equal to $(3/T_c)^3 \approx (2.2{\rm fm})^3$ 
in physical units:
\begin{equation}
T_c/\sqrt{\sigma}({\rm finite\ volume}) = \left\{
\begin{array}{ll}
0.660(3)(^{+14}_{-10}) & (N_t=3) \\
0.647(5)          & (N_t=4) \\
0.651(6)          & (N_t=6)  {_.}
\end{array}
\right.
\label{eqn:Tcsfv}
\end{equation}
The number in the first brackets is the statistical error
and the second one for $N_t=3$ is the systematic error
due to uncertainty of the fitting range.

In order to estimate the values of $T_c/\sqrt{\sigma}$ 
in the infinite volume limit,
we first obtain finite size scaling relations \cite{fukugita,StdNt46}
\begin{equation}
\beta_c(N_t,N_s^3) =
\beta_c(N_t,\infty) - 
{0.122(54)}\;{\cdot}\;{{N_t}^3}/{N_s^3} \;\;(N_t=3),
\end{equation}
and
\begin{equation}
\beta_c(N_t,N_s^3) =
\beta_c(N_t,\infty) - 
{0.133(63)}\;{\cdot}\;{{N_t}^3}/{N_s^3} \;\;(N_t=4),
\label{eqn:FSSNt4}
\end{equation}
from the data of $\beta_c$ on the 
$N_s/N_t=3$, 4, and 5 lattices (see Fig.~\ref{fig:FSSbc}).
We note that the slopes of $\beta_c(N_t,N_s^3)$ in $N_t^3/N_s^3$ 
in the two relations are independent of $N_t$ within the errors, 
as observed previously in the case of 
the standard one-plaquette action \cite{StdNt46}.
Therefore, we assume the relation (\ref{eqn:FSSNt4}) also 
for $N_t=6$. 
Then we have 
\begin{equation}
\beta_c(N_t,\infty) = \left\{
\begin{array}{ll}
2.1551(12) & (N_t=3) \\
2.2879(11) & (N_t=4) \\
2.5206(30) & (N_t=6)  {_.}
\end{array}
\right.
\end{equation}

The values of the string tension at $\beta_c(N_t,\infty)$
are estimated assuming an exponential scaling of 
$\sqrt{\sigma_{\rm lat}}$ in
terms of $\beta$ \cite{StdTcs2}.
We obtain 
\begin{equation}
\sqrt{\sigma_{\rm lat}} = 25.26(43)\;{\cdot}\; 
\exp \left[ \; -1.828(7) {\cdot} \beta  \;\right],
\end{equation}
by fitting the values of $\sigma_{\rm lat}$ 
at $\beta_c(9{^3}\times3)$, 
$\beta_c(12{^3}\times4)$, and $\beta_c(18{^3}\times6)$ 
as shown in Fig.~\ref{fig:Sigma}.
This relation is used to compute the shifts in $\sigma_{\rm lat}$ 
from the values at $\beta_c(N_t,N_s^3)$ to those at 
$\beta_c(N_t,\infty)$.
The values of $\sigma_{\rm lat}(\beta_c(N_t,\infty))$ are
obtained by adding the shifts to 
those of $\sigma_{\rm lat}(\beta_c(N_t,N_s^3))$ 
given in Tables~\ref{tab:FitRst} and \ref{tab:FitRstNt3}:
\begin{equation}
\sigma_{\rm lat}(\beta_c(N_t,\infty)) = \left\{
\begin{array}{ll}
0.2516(26)(11)(^{+72}_{-110})   & (N_t=3) \\
0.1464(25)(6)            & (N_t=4) \\
0.0644(12)(7)            & (N_t=6)  {_.}
\end{array}
\right.
\label{eqn:Sigmas}
\end{equation}
The number in the first brackets is the statistical error,
the second one is the error due to the error in the values of 
$\beta_c(N_t,\infty)$,
and the third one for $N_t=3$ is the systematic error
due to uncertainty of the fitting range.

Finally, we obtain 
\begin{equation}
T_c/\sqrt{\sigma}({\rm infinite\ volume}) = \left\{
\begin{array}{ll}
0.665(3)(1)(^{+15}_{-10})   & (N_t=3) \\
0.653(6)(1)            & (N_t=4) \\
0.657(6)(4)            & (N_t=6)  {_.}
\end{array}
\right.
\label{eqn:Tcs}
\end{equation}
The origins of the errors are the same as in Eq.(\ref{eqn:Sigmas}).
Our three values are consistent with a constant within the errors.
A weighted average of the values given in Eq.(\ref{eqn:Tcs}) gives
\begin{equation}
T_c/\sqrt{\sigma} = 0.656(4)
\label{eqn:extrTcs}
\end{equation}
in the continuum limit.

Using the experimental value 
$\sigma=\sigma_{\rm lat}/a^2=(420\mbox{MeV})^2$,
we obtain $a \approx 0.23$, 0.18, and 0.12 fm
at $\beta_c$ for $N_t=3$, 4, and 6, respectively.
Thus the scaling behavior for the ratio $T_c/\sqrt{\sigma}$
starts at least around  $a \approx 0.23$fm 
with the RG improved gauge action.
From Eq.(\ref{eqn:extrTcs}) 
we also obtain $T_c\approx 276(2)\mbox{MeV}$.

Our results ({\ref{eqn:Tcs}})
are shown in Fig.~\ref{fig:Tcs} together with 
the results using other actions \cite{BSATcs,StdTcs2}.
Our result $T_c/\sqrt{\sigma} = 0.656(4)$ in the continuum limit
is slightly larger than the value
with the standard action $0.629(3)$ \cite{StdTcs2}.
We also compare our results with those derived from 
the torelon mass $\mu(L)$ which is calculated from Polyakov line 
correlators
on a lattice of spatial size $L$.
Defining $\sigma(L)=\mu(L)/L$, we extrapolate 
the values of $T_c/\sqrt{\sigma(L)}$ to the continuum limit.
Then the value of $T_c/\sqrt{\sigma}$ is estimated 
assuming the relation $\sigma=\sigma(L)+\pi/(3L^2)$ derived
in a string model \cite{PhdeF}.
(We neglect the corrections due to the shift 
$\beta_c(V=\infty)-\beta_c(V=N_s^3)$.)
For a fixed point action \cite{FA}, we obtain  
$T_c/\sqrt{\sigma}$ = 0.617(5) using the data for $N_t=2$, 3 and 4 
with $N_s=2N_t$.
The result is about 6\% smaller than our result (\ref{eqn:extrTcs}).
For a tadpole-improved Symanzik action \cite{SLW},
we obtain $T_c/\sqrt{\sigma}$ = 0.649(5) using the data for 
$N_t = 3$ and 4 with $N_s=2N_t$.
The result is consistent with our result.

Numerical simulations are performed with Fujitsu VPP500/30 and
HITAC H6080-FP12 at the University of Tsukuba.
We thank Akira Ukawa for valuable discussions.
This work is in part supported by 
the Grants-in-Aid of Ministry of Education,
Science and Culture (Nos.07NP0401, 07640375 and 07640376)
and the University of Tsukuba Project Research in 1996.

\clearpage

\begin{table} [t]
\vspace{10mm}
  \caption{Parameters of finite temperature simulations}
  \label{tab:FTPTpara}
\begin{center}
  \begin{tabular}{clrr}                   \hline
    \makebox[30mm][c]{lattice size}   &
    \makebox[20mm][l]{$\beta$}        &  
    \makebox[20mm][r]{sweep}          & 
    \makebox[20mm][r]{therm.} \\ \hline
      $9^3\times3$  &  2.125  &  100000 & 30000  \\
                    &  2.145  &  100000 & 30000  \\ 
                    &  2.160  &   70000 & 30000  \\ \hline
     $12^3\times3$  &  2.150  &  100000 & 50000  \\ 
                    &  2.155  &  100000 & 40000  \\ \hline
     $15^3\times3$  &  2.150  &  180000 & 80000  \\ \hline
     $12^3\times4$  &  2.250  &  12000  &   2000  \\
                    &  2.275  &  125000 &  40000  \\
		    &  2.300  &  10000  &   1500  \\ \hline
     $16^3\times4$  &  2.283  &  220000 &  40000  \\
		    &  2.290  &  240000 &  40000  \\ \hline
     $20^3\times4$  &  2.2875 &  270000 &  80000  \\ \hline
     $18^3\times6$  &  2.5000  &  120000 &  15000  \\
                    &  2.5125  &  256000 &  50000  \\
                    &  2.5250  &  210000 &  60000  \\ 
                    &  2.5375  &  135000 &   5000  \\ \hline
\end{tabular}
\end{center}
\vspace{-05mm}
\end{table}

\begin{table} [b]
    \vspace{10mm}
    \caption{Results obtained on the $9{^3}\times3$ lattice.}
    \label{tab:FTPTmeas1}
   \vspace{-5mm}
\begin{center}
  \begin{tabular}{clll} \hline
    \makebox[40mm][c]{} &
    \makebox[30mm][l]{$\beta=2.125$} &  
    \makebox[30mm][l]{$\beta=2.145$} &
    \makebox[30mm][l]{$\beta=2.160$} \\ \hline
     $1\times1$ Wilson loop  &   
          0.575350(61)  &   0.58207(12)  &
          0.58752(15)   \\ 
     $1\times2$ Wilson loop  &    
          0.32181(11)   &   0.33120(22)  &
          0.33917(28)   \\   
     $2\times2$ Wilson loop  &    
          0.10721(18)   &   0.11600(32)  &
          0.12436(43)   \\   
     Polyakov line    &        
          0.0675(23)    &   0.1055(40)   &
          0.1575(52)    \\   
     deconfinement fraction    &        
          0.448(23)     &   0.666(29)    &
          0.895(29)     \\   \hline
\end{tabular}
\end{center}
\vspace{-5mm}
\end{table}

\begin{table} [b]
    \vspace{5mm}
    \caption{Results obtained on the $12{^3}\times3$ lattice.}
\begin{center}
  \begin{tabular}{cll} \hline
    \makebox[40mm][c]{} &
    \makebox[30mm][l]{$\beta=2.150$} &  
    \makebox[30mm][l]{$\beta=2.155$} \\ \hline
     $1\times1$ Wilson loop  &   
          0.58329(16)   &    0.58546(11)  \\ 
     $1\times2$ Wilson loop  &    
          0.33277(31)   &    0.33607(20)  \\   
     $2\times2$ Wilson loop  &    
          0.11716(47)   &    0.12094(31)  \\   
     Polyakov line      &         
          0.0972(74)    &    0.1300(38)   \\ 
     deconfinement fraction    &         
          0.691(53)     &    0.899(24)    \\ \hline  
\end{tabular}
\end{center}
\vspace{5mm}
\end{table}

\begin{table} [b]
    \vspace{5mm}
    \caption{Results obtained on the $15{^3}\times3$ lattice.}
\begin{center}
  \begin{tabular}{cl} \hline
    \makebox[40mm][c]{} &
    \makebox[30mm][l]{$\beta=2.150$} \\ \hline
     $1\times1$ Wilson loop  &   0.58321(17)  \\ 
     $1\times2$ Wilson loop  &   0.33260(32)  \\   
     $2\times2$ Wilson loop  &   0.11688(48)  \\   
     Polyakov line           &   0.0862(57)   \\
     deconfinement fraction  &   0.715(52)    \\ \hline  
\end{tabular}
\end{center} 
\vspace{5mm}
\end{table}

\begin{table} [b]
    \vspace{5mm}
    \caption{Results obtained on the $12{^3}\times4$ lattice.}
   \vspace{-5mm}
\begin{center}
  \begin{tabular}{clll} \hline
    \makebox[40mm][c]{} &
    \makebox[30mm][l]{$\beta=2.250$} &  
    \makebox[30mm][l]{$\beta=2.275$} & 
    \makebox[30mm][l]{$\beta=2.300$} \\ \hline
     $1\times1$ Wilson loop  &  0.608085(91)   &
       0.614037(60)   &  0.62007(28)  \\ 
     $1\times2$ Wilson loop  &  0.36552(18)  &
       0.37384(12)  &  0.38255(52)  \\   
     $2\times2$ Wilson loop  &  0.14448(25)  &
       0.15257(19)  &  0.16175(76)  \\ 
     Polyakov line    &  0.0374(25)   &  
       0.0651(31)   &  0.1213(16)   \\ 
     deconfinement fraction    &  0.342(33)   &  
        0.640(23)  &  0.981(23)   \\ \hline  
\end{tabular}
\end{center}
\vspace{5mm}
\end{table}

\begin{table} [t]
    \caption{Results obtained on the $16{^3}\times4$ lattice.}
\begin{center}
  \begin{tabular}{cll} \hline
    \makebox[40mm][c]{} &
    \makebox[30mm][l]{$\beta=2.283$} &  
    \makebox[30mm][l]{$\beta=2.290$} \\ \hline
     $1\times1$ Wilson loop  &  0.615677(33)   &
       0.617594(35)   \\ 
     $1\times2$ Wilson loop  &  0.376040(65)  &
       0.378940(69)  \\   
     $2\times2$ Wilson loop  &  0.15451(10)  &
       0.15781(11))  \\  
     Polyakov line    &  0.0549(30)   &  
       0.0906(30)   \\   
     deconfinement fraction  &  0.583(39)   &  
       0.853(27)    \\   \hline
\end{tabular}
\end{center}
\end{table}

\begin{table} [b]
    \vspace{5mm}
    \caption{Results obtained on the $20{^3}\times4$ lattice.}
\begin{center}
  \begin{tabular}{cl} \hline
    \makebox[40mm][c]{} &
    \makebox[30mm][l]{$\beta=2.2875$} \\ \hline
     $1\times1$ Wilson loop  &   0.616941(56)  \\ 
     $1\times2$ Wilson loop  &   0.37794(11)   \\   
     $2\times2$ Wilson loop  &   0.15663(19)  \\   
     Polyakov line           &   0.0744(57)   \\ 
     deconfinement fraction  &   0.768(48)    \\ \hline  
\end{tabular}
\end{center} 
\vspace{5mm}
\end{table}

\begin{table} [b]
   \caption{Results obtained on the $18{^3}\times6$ lattice.}
   \label{tab:FTPTmeas5}
   \vspace{-5mm}
\begin{center}
  \begin{tabular}{cllll} \hline
    \makebox[35mm][c]{} &
    \makebox[25mm][l]{$\beta=2.5000$} &  
    \makebox[25mm][l]{$\beta=2.5125$} &
    \makebox[25mm][l]{$\beta=2.5250$} &
    \makebox[25mm][l]{$\beta=2.5375$} \\ \hline
     $1\times1$ Wilson loop  &  0.655687(59)   &  0.657691(11) &
       0.659676(11)   &  0.661649(12)  \\ 
     $1\times2$ Wilson loop  &  0.431669(40)   &  0.434539(25) &
       0.437385(24)   &  0.440230(24)  \\   
     $1\times3$ Wilson loop  &  0.288927(53)   &  0.291912(33) &
       0.294884(33)   &  0.297874(32)  \\   
     $2\times2$ Wilson loop  &  0.208714(77)   &  0.211717(46) &
       0.214735(47)   &  0.217780(34)  \\   
     $2\times3$ Wilson loop  &  0.109884(87)   &  0.112330(53) &
       0.114813(54)   &  0.117350(55)  \\   
     $3\times3$ Wilson loop  &  0.049770(87)   &  0.051480(54) &
       0.053261(58)   &  0.055096(59)  \\   
     Polyakov line    &  0.0328(32)   &  0.0409(21) &
       0.0559(22)   &  0.0691(21)  \\ 
     deconfinement fraction    &  0.555(51)   &  0.645(36) &
       0.861(31)    &  0.960(15) \\ \hline  
\end{tabular}
\end{center}
\vspace{5mm}
\end{table}

\begin{table} [t]
   \vspace{10mm}
   \caption{The value of $\beta{_c}$ and its jack knife error 
            determined by the peak location of 
            the susceptibility, and
            the deconfinement fraction $r$ at $\beta_c$,
            together with the bin size adopted.}
   \label{tab:betac}
\begin{center}
  \begin{tabular}{clll} \hline
    \makebox[30mm][c]{} &
    \makebox[30mm][l]{$\beta_c$              } &  
    \makebox[30mm][l]{$r$                    } &  
    \makebox[20mm][l]{bin size               } \\ \hline
        \makebox[30mm][c]{ $9{^3}\times3$} & 2.1508(12) &
                                 0.757(25) &       1000 \\
        \makebox[30mm][c]{$12{^3}\times3$} & 2.1528(9)  & 
                                 0.771(48) &       3000 \\
        \makebox[30mm][c]{$15{^3}\times3$} & 2.1546(11) & 
                                 0.894(33) &       8000 \\ \hline
        \makebox[30mm][c]{$12{^3}\times4$} & 2.2827(16) &
                                 0.774(47) &       3000 \\
        \makebox[30mm][c]{$16{^3}\times4$} & 2.2863(10) & 
                                 0.765(37) &       6000 \\
        \makebox[30mm][c]{$20{^3}\times4$} & 2.2865(9)  & 
                                 0.742(52) &      10000 \\ \hline
        \makebox[30mm][c]{$18{^3}\times6$} & 2.5157(7)  & 
                                 0.698(34) &      3000  \\ \hline
\end{tabular}
\end{center}
\vspace{10mm}
\end{table}

\begin{table} [tb]
   \vspace{10mm}
   \caption{Deconfinement fraction $r$ at $\beta{_c}$ determined 
            from the susceptibility 
            in the case of the standard one-plaquette action. 
            Data of the Polyakov line are taken from a previous 
            simulation by the 
            QCDPAX Collaboration \protect\cite{StdNt46}.}
   \label{tab:stdr}
\begin{center}
  \begin{tabular}{cll} \hline
    \makebox[50mm][c]{lattice} &
    \makebox[30mm][l]{$\beta_c$} &  
    \makebox[30mm][l]{$r$           } \\ \hline
        \makebox[30mm][c]{$12^3\times24\times4$} & 5.69149  & 
                               0.790(12) \\
        \makebox[30mm][c]{$24^3\times36\times4$} & 5.69245  & 
                               0.732(46) \\ \hline
        \makebox[30mm][c]{$20{^3}\times6$} & 5.8924 & 
                               0.805(26) \\
        \makebox[30mm][c]{$24{^3}\times6$} & 5.89292 & 
                               0.786(27) \\
        \makebox[30mm][c]{$36^3\times48\times6$} & 5.89379 & 
                               0.739(40) \\ \hline
\end{tabular}
\end{center}
\vspace{10mm}
\end{table}

\clearpage

\begin{table}[t]
\vspace{20mm}
    \caption{Simulation parameters for the quark potential
             measurements.}
    \label{tbl:STpara}
\begin{center}  
\begin{tabular}{cccc} \hline
    \makebox[30mm][c]{lattice size}    & 
    \makebox[20mm][c]{$\beta$}         &
    \makebox[30mm][c]{thermalization}  & 
    \makebox[20mm][c]{{\#} of conf.} \\  
\hline
     $9^3\times 18$  &  2.1508  &   5000  &  400  \\ 
    $12^3\times 24$  &  2.2827  &   5000  &  200  \\ 
    $18^3\times 36$  &  2.5157  &  10000  &  100  \\  \hline
\end{tabular}
\end{center}
    \label{tab:STpara}
\vspace{0mm}
\end{table}

\begin{table} [t]
    \caption{Potential $V(R)$ and overlap function $C(R)$ 
             at $\beta_c(12{^3}{\times}4)$ 
             obtained on the $12{^3}\times24$ lattice.
             The optimum smearing step $N_{\rm opt}$ is also given.
             }
    \label{tab:STmeas1}
\begin{center}
\begin{tabular}{llllr} \hline
    \makebox[20mm][l]{$R$} &
    \makebox[20mm][l]{unit of R} &  
    \makebox[30mm][l]{$V(R)$} &
    \makebox[30mm][l]{$C(R)$} &
    \makebox[10mm][r]{$N_{opt}$} \\ \hline
    1.000        &  (1,0,0)     &
    0.47408(35)  &  0.9997(8)   &
    10  \\ 
    1.414        &  (1,1,0)     &
    0.63207(60)  &  0.9987(14)  &
    10  \\  
    1.732        &  (1,1,1)     &
    0.72136(89)  &  0.9955(23)  &
    10  \\  
    2.000        &  (1,0,0)     &
    0.7811(12) &  0.9916(31)  &
    12  \\  
    2.236        &  (2,1,0)     &
    0.8304(11) &  0.9880(25)  &
    10  \\  
    2.449        &  (2,1,1)     &
    0.8759(12) &  0.9977(29)  &
     8  \\  
    2.828        &  (1,1,0)     &
    0.9472(17) &  0.9990(43)  &
     8  \\  
    3.000        &  (1,0,0)     &
    0.9843(27) &  0.9857(72)  &
    10  \\  
    3.000        &  (2,2,1)     &
    0.9806(19) &  0.9799(48)  &
     8  \\  
    3.464        &  (1,1,1)     &
    1.0574(32) &  0.9739(86)  &
     8  \\  
    4.000        &  (1,0,0)     &
    1.1507(15) &  0.9779(41)  &
     8  \\  
    4.243        &  (1,1,0)     &
    1.1926(45) &  0.982(12) &
     8  \\  
    4.472        &  (2,1,0)     &
    1.2317(41) &  0.976(11) &
     8  \\  
    4.899        &  (2,1,1)     &
    1.3053(55) &  0.971(15) &
     8  \\  
    5.000        &  (1,0,0)     &
    1.3222(22) &  0.9881(63)  &
     8  \\  
    5.196        &  (1,1,1)     &
    1.34430(55)  &  0.9925(57)  &
     6  \\  
    5.657        &  (1,1,0)     &
    1.426(11)&  0.982(29) &
     8  \\  
    6.000        &  (1,0,0)     &
    1.479(19)&  0.978(53) &
     8  \\  
    6.000        &  (2,2,1)     &
    1.49012(78)  &  0.9966(66)  &
     6  \\ \hline
\end{tabular}
\end{center}
\vspace{-05mm}
\end{table}

\clearpage

\begin{table} [t]
    \caption{The same as Table~\protect\ref{tab:STmeas1} 
             at $\beta_c(18{^3}{\times}6)$
             obtained on the $18{^3}\times36$ lattice.}
    \label{tab:STmeas2}
\begin{center}
\begin{tabular}{llllr} \hline
    \makebox[20mm][l]{$R$} &
    \makebox[20mm][l]{unit of R} &  
    \makebox[30mm][l]{$V(R)$} &
    \makebox[30mm][l]{$C(R)$} &
    \makebox[10mm][r]{$N_{opt}$} \\ \hline
   1.000        &  (1,0,0)     &
    0.39187(22)  &  1.0059(7)   &
    40  \\  
    1.414        &  (1,1,0)     &
    0.50809(29)  &  1.0065(10)  &
    40  \\  
    1.732        &  (1,1,1)     &
    0.56917(44)  &  1.0059(17)  &
    40  \\  
    2.000        &  (1,0,0)     &
    0.60693(62)  &  1.0042(21)  &
    40  \\  
    2.236        &  (2,1,0)  &
    0.63812(56)  &  0.9991(19)  &
    36  \\  
    2.449        &  (2,1,1)  &
    0.66409(68)  &  0.9991(21)  &
    30  \\  
    2.828        &  (1,1,0)  &
    0.70597(77)  &  0.9964(25)  &
    28  \\  
    3.000        &  (1,0,0)  &
    0.7270(12) &  0.9983(40)  &
    30  \\  
    3.000        &  (2,2,1)  &
    0.72320(82)  &  0.9986(27)  &
    24  \\  
    3.464        &  (1,1,1)  &
    0.7671(14) &  0.9970(49)  &
    22  \\  
    4.000        &  (1,0,0)  &
    0.8151(18) &  0.9985(67)  &
    20  \\  
    4.243        &  (1,1,0)  &
    0.8340(18) &  0.9915(58)  &
    20  \\  
    4.472        &  (2,1,0)  &
    0.8548(17) &  0.9933(56)  &
    20  \\  
    4.899        &  (2,1,1)  &
    0.8872(19) &  0.9914(65)  &
    18  \\  
    5.000        &  (1,0,0)  &
    0.8929(31) &  1.000(11) &
    16  \\  
    5.196        &  (1,1,1)  &
    0.9089(23) &  1.0000(90)  &
    16  \\  
    5.657        &  (1,1,0)  &
    0.9462(28) &  0.9894(97)  &
    18  \\  
    6.000        &  (1,0,0)  &
    0.9644(39) &  0.993(15) &
    14  \\  
    6.000        &  (2,2,1)  &
    0.9721(29) &  0.992(10) &
    16  \\  
    6.708        &  (2,1,0)  &
    1.0254(34) &  0.982(11) &
    18  \\  
    6.928        &  (1,1,1)  &
    1.0374(43) &  0.984(15) &
    16  \\  
    7.000        &  (1,0,0)  &
    1.04334(53)  &  0.9905(64)  &
    16  \\  
    7.071        &  (1,1,0)  &
    1.0501(18) &  0.9983(84)  &
    16  \\  
    7.348        &  (2,1,1)  &
    1.0674(41) &  0.995(14) &
    14  \\  
    8.000        &  (1,0,0)  &
    1.11216(68)  &  0.9953(78)  &
    14  \\  
    8.485        &  (1,1,0)  &
    1.15463(52)  &  0.9998(84)  &
    16  \\  
    8.660        &  (1,1,1)  &
    1.1714(32) &  0.988(14) &
    18  \\  
    8.944        &  (2,1,0)  &
    1.1743(22) &  0.9870(78)  &
    12  \\  
    9.000        &  (1,0,0)  &
    1.17869(54)  &  0.9794(86)  &
    14  \\  
    9.000        &  (2,2,1)  &
    1.18690(47)  &  0.9975(78)  &
    12  \\   \hline
\end{tabular}
\end{center}
\vspace{-05mm}
\end{table}

\clearpage

\begin{table}[t]
    \caption{Results of $V_0$, $\alpha$, and $\sigma_{\rm lat}$
             at $\beta_c(12{^3}{\times}4)$ 
             and $\beta_c(18{^3}{\times}6)$
             obtained on the $12{^3}{\times}24$ 
             and $18{^3}{\times}36$ lattices, respectively.}
    \label{tab:FitRst}
\vspace{-0.6cm}
\begin{center}  
\begin{tabular}{cccc} \hline
\makebox[30mm][c]{$\beta$}  & \makebox[40mm][c]{$V_0$}    &
\makebox[20mm][c]{$\alpha$} & \makebox[40mm][c]{$\sigma_{\rm lat}$} \\ 
\hline
    $\beta_c(12{^3}\times4)$  &  0.630(20)  &
                   0.295(14)  & 0.1493(25)  \\
    $\beta_c(18{^3}\times6)$  &  0.627(18)  &
                   0.297(19)  & 0.0655(12)  \\  \hline
\end{tabular}
\end{center}
\vspace{-5mm}
\end{table}

\begin{table} [b]
    \caption{The same as Table~\protect\ref{tab:STmeas1} 
             at $\beta_c(9^3 \times 3)$ 
             obtained on the $9{^3}\times18$ lattice.}
    \label{tab:STmeas3}
\begin{center}
\begin{tabular}{llllc} \hline
    \makebox[20mm][l]{$R$} &
    \makebox[20mm][l]{unit of R} &  
    \makebox[30mm][l]{$V(R)$} &
    \makebox[30mm][l]{$C(R)$} &
    \makebox[10mm][c]{$N_{opt}$} \\ \hline
    1.000  &  (1,0,0)  &
    0.54892(35)  &  0.9982(5)  &      4  \\ 
    1.414  &  (1,1,0)  &
    0.75390(63)  &  0.9984(10)  &      4  \\
    1.732  &  (1,1,1)  &
    0.87752(99)  &  0.9942(16)  &      4  \\
    2.000  &  (1,0,0)  &
    0.9601(13)  &  0.9981(22)  &      5  \\
    2.236  &  (2,1,0)  &
    1.0349(13)  &  0.9965(21)  &      4  \\
    2.449  &  (2,1,1)  &
    1.1049(15)  &  0.9963(25)  &      2  \\
    2.828  &  (1,1,0)  &
    1.2101(22)  &  0.9945(39)  &      4  \\
    3.000  &  (1,0,0)  &
    1.2640(31)  &  0.9836(55)  &      5  \\
    3.000  &  (2,2,1)  &
    1.2685(23)  &  0.9591(39)  &      3  \\
    3.464  &  (1,1,1)  &
    1.3890(41)  &  0.9736(73)  &      4  \\
    4.000  &  (1,0,0)  &
    1.5420(18)  &  0.9750(34)  &      5  \\
    4.243  &  (1,1,0)  &
    1.6001(60)  &  0.987(12)  &      4  \\
    4.472  &  (2,1,0)  &
    1.6634(62)  &  0.978(11)  &      4  \\
    4.899  &  (2,1,1)  &
    1.7785(53)  &  0.973(13)  &      3  \\ \hline
\end{tabular}
\end{center}
\vspace{-0mm}
\end{table}

\begin{table}[t]
    \caption{Results of $V_0$ and $\sigma_{\rm lat}$ 
             at $\beta_c(9^3 \times 3)$ 
             obtained on the $9^3 \times 18$ lattice.
             The numbers in the first brackets are statistical 
errors and the second are systematic errors due to uncertainty 
of the fitting range.}
    \label{tab:FitRstNt3}
\vspace{0mm}
\begin{center}  
\begin{tabular}{cc} \hline
    \makebox[40mm][c]{$V_0$}    &
    \makebox[40mm][c]{$\sigma_{\rm lat}$} \\ 
\hline
    0.598(60)($^{+57}_{-48}$)  &  0.2554(26)($^{+72}_{-110}$) \\ 
\hline
\end{tabular}
\end{center}
\vspace{0.5cm}
\end{table}

\clearpage

\begin{figure}[p]
\begin{center}
    \leavevmode
    \epsfxsize=8.5cm 
    \epsfbox{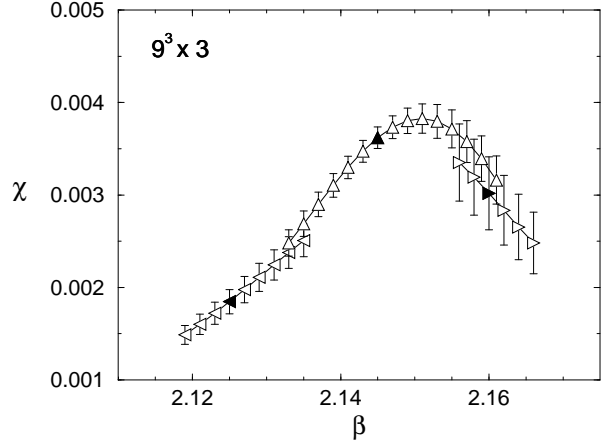}
\vspace{10mm}
    \epsfxsize=8.5cm 
    \epsfbox{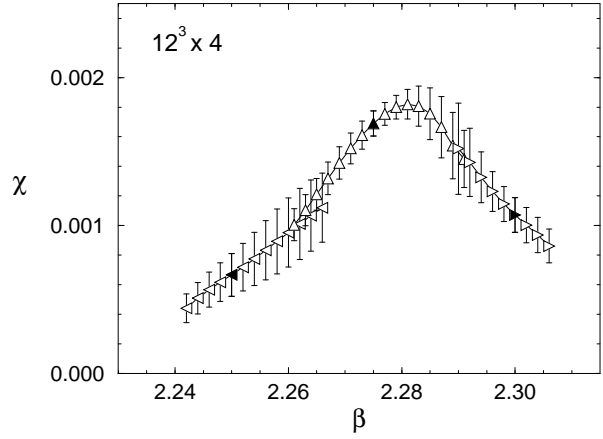}
\end{center}
\vspace{0mm}
    \caption{Susceptibility $\chi$
of the Z(3) rotated Polyakov line $\Omega$
on the $9{^3}{\times}3$,
             $12{^3}{\times}4$ and 
             $18{^3}{\times}6$ lattices.
	     Each curve is obtained from the data
	     at the simulation point indicated by
             the filled symbol using the spectral 
             density method.
             Errors for open symbols are computed 
             by a jack-knife method combined with the 
             spectral density method.}
\label{fig:suscep}
\vspace{-2mm}
\end{figure}

\setcounter{figure}{0}

\begin{figure}[p]
\begin{center}
    \leavevmode
    \epsfxsize=8.5cm 
    \epsfbox{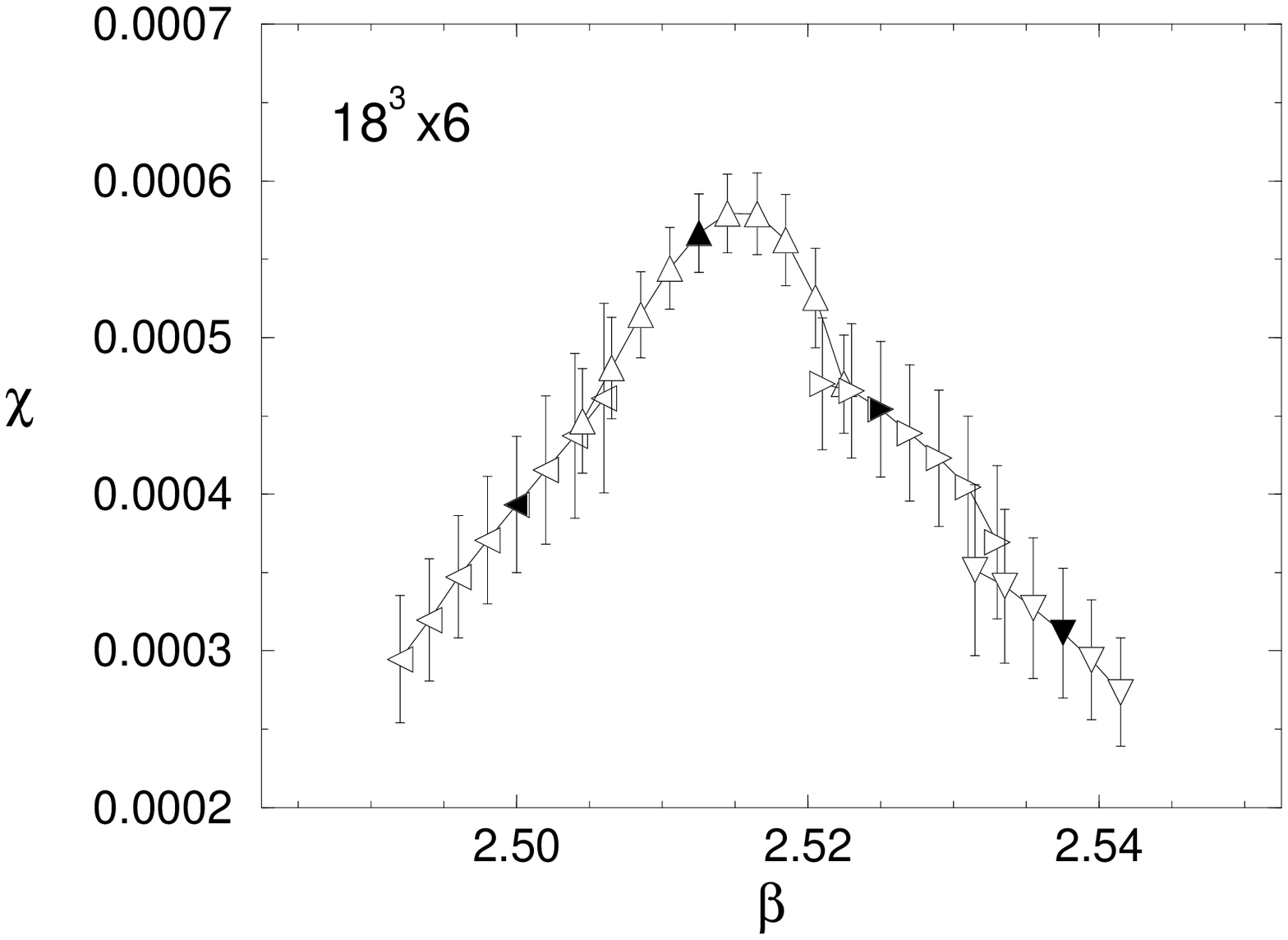}
\end{center}
    \caption{\it continued}
\end{figure}

\clearpage

\begin{figure}[p]
    \vspace{-5mm}
\begin{center}
    \leavevmode
    \epsfxsize=9cm 
    \epsfbox{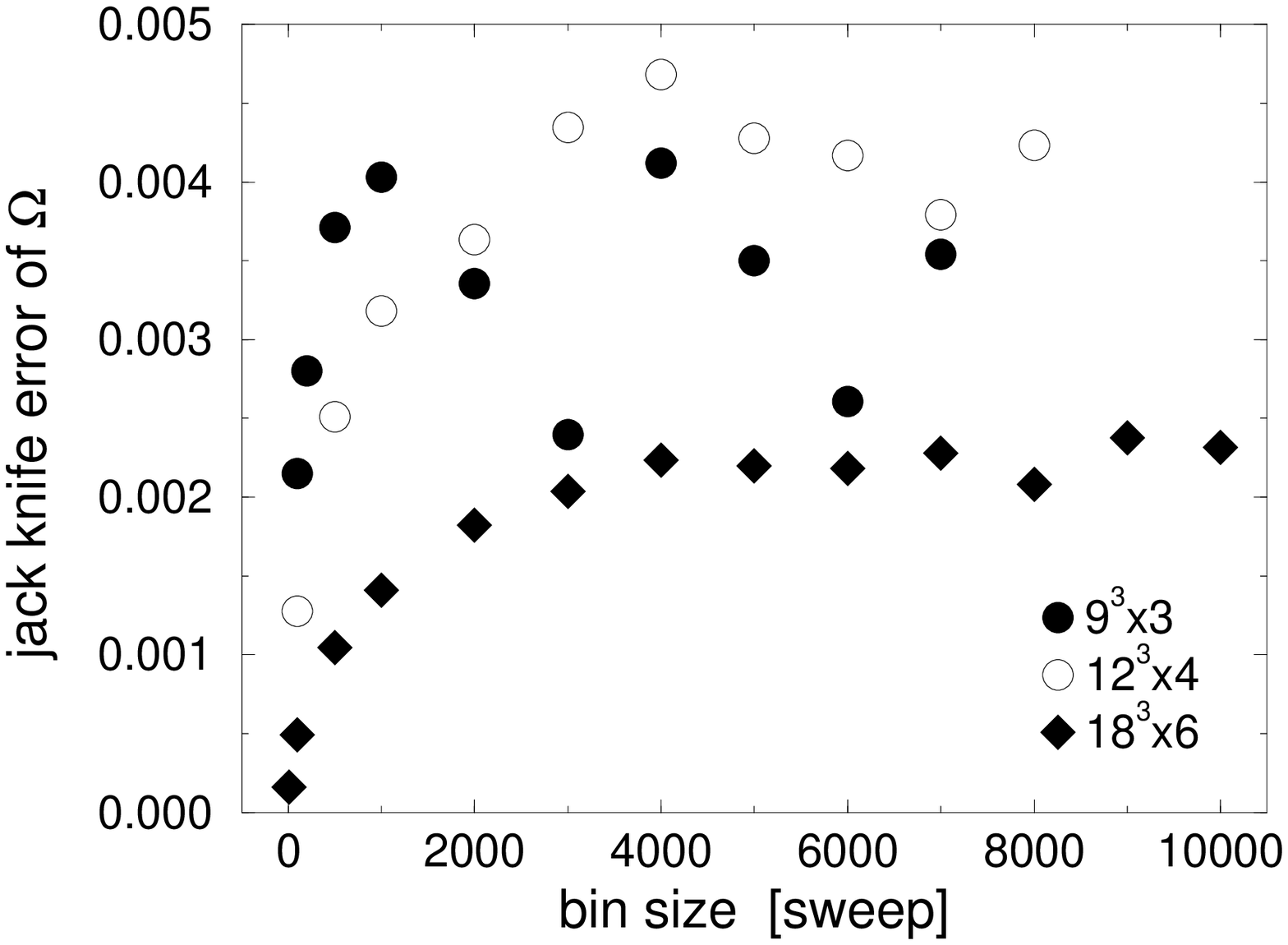}
\end{center}
\vspace{-0.5cm}
    \caption{Bin size dependence of the jack knife error of $\Omega$.}
\label{fig:pjne}
\vspace{-5mm}
\end{figure}

\begin{figure}[p]
\begin{center}
    \leavevmode
    \epsfxsize=9cm 
    \epsfbox{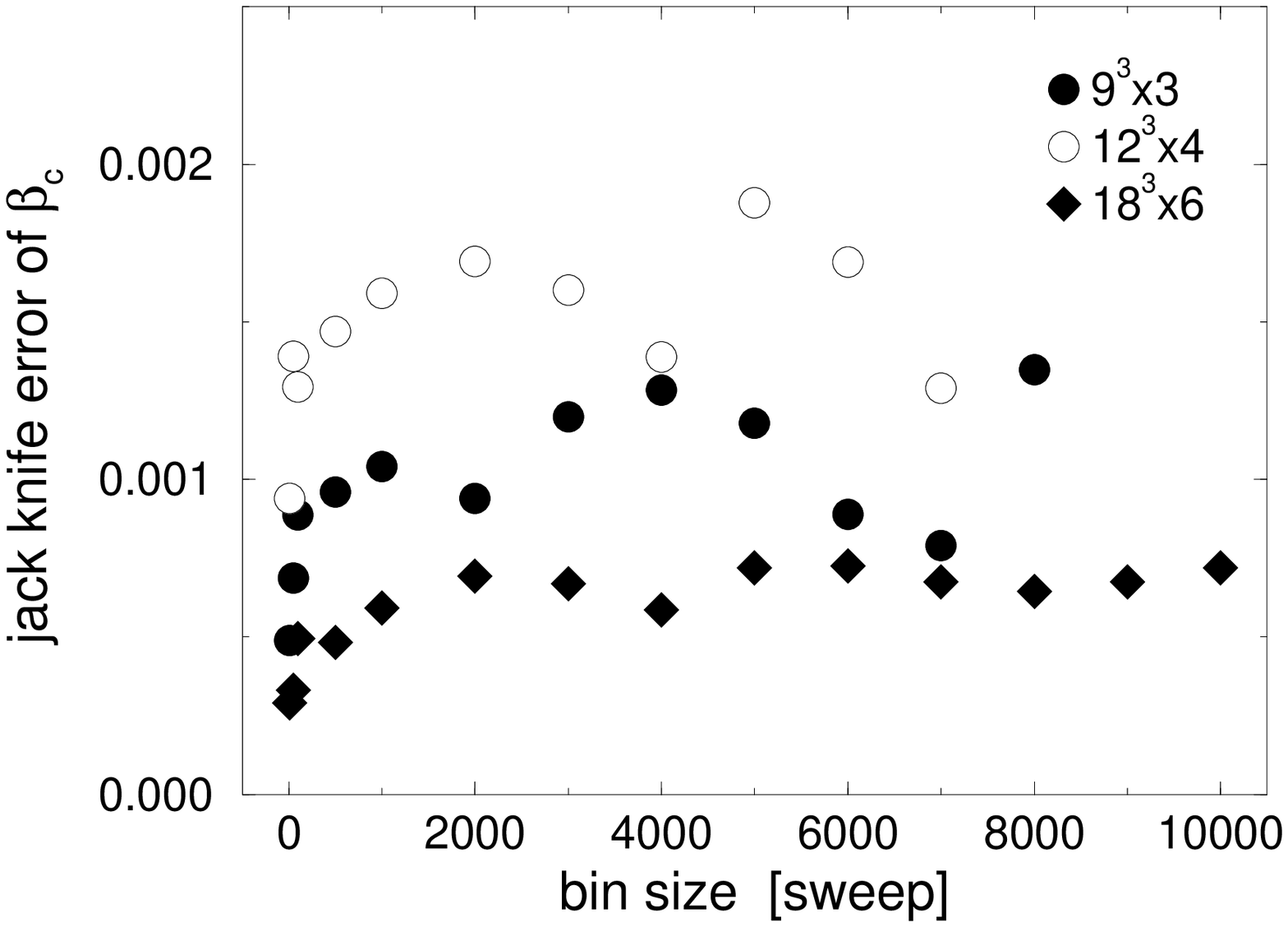}
\end{center}
\vspace{-0.5cm}
    \caption{Bin size dependence of the jack knife error of $\beta_c$.}
\label{fig:bcjne}
\vspace{0.4cm}
\end{figure}

\clearpage

\begin{figure}[p]
\begin{center}
    \leavevmode
    \epsfxsize=9.0cm 
    \epsfbox{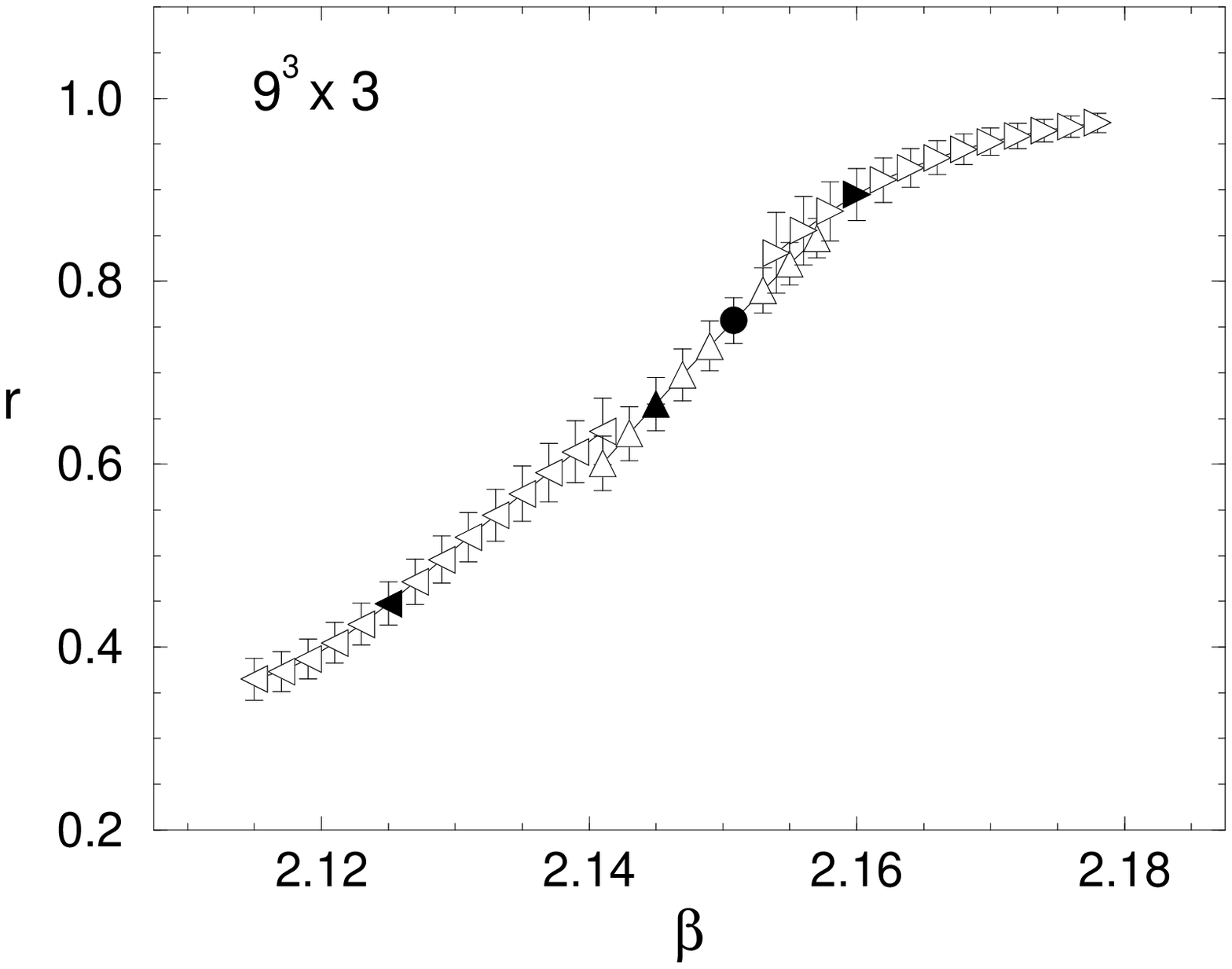}
\vspace{10mm}
    \epsfxsize=9.0cm 
    \epsfbox{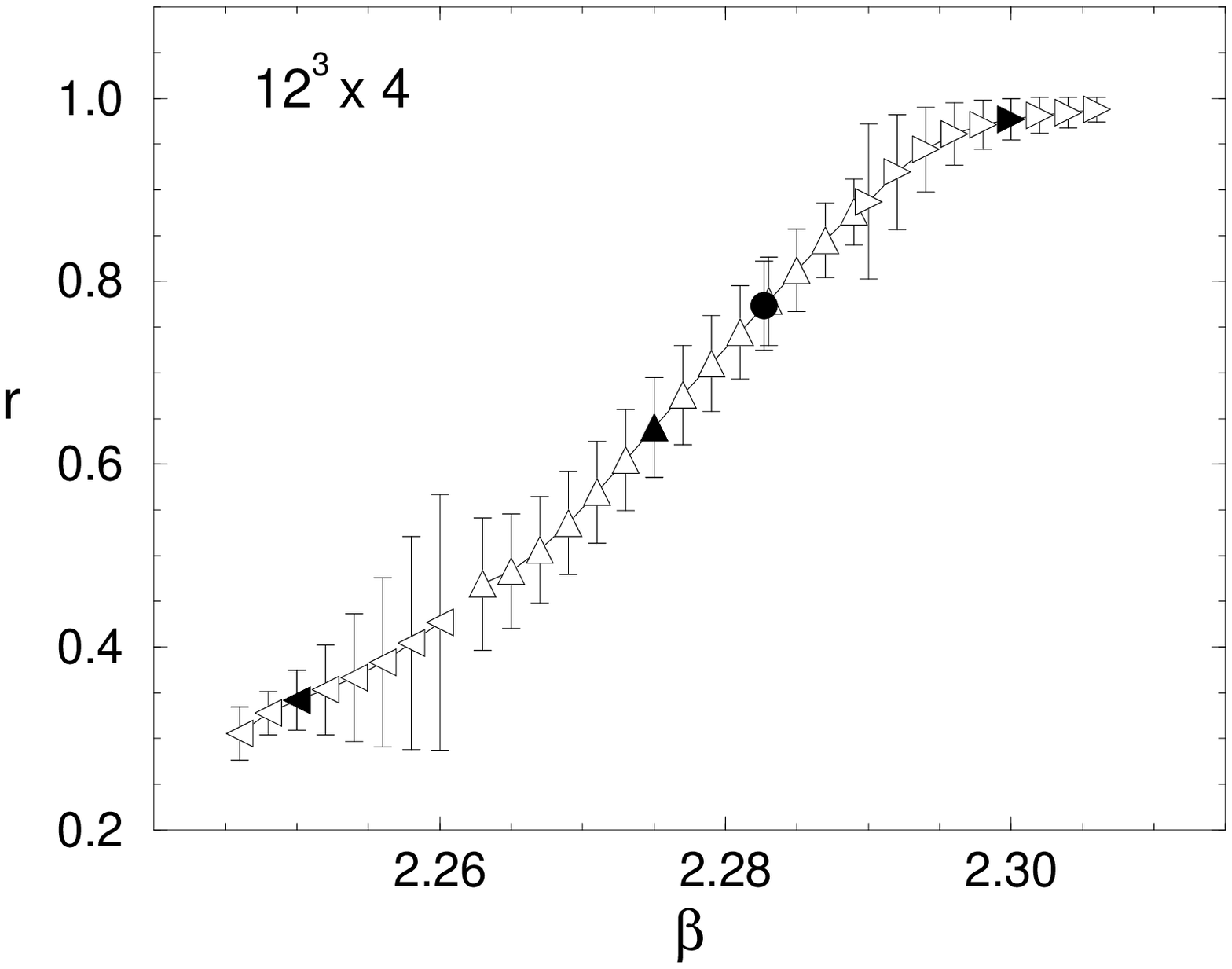}
\end{center}
\vspace{0mm}
    \caption{Deconfinement fraction $r$
             on the $9{^3}{\times}3$,
                    $12{^3}{\times}4$ and 
                    $18{^3}{\times}6$ lattices.
	     Each curve is obtained from the data
	     at the simulation point indicated by
             filled triangles.
             The values of $\beta_c$ determined from the peak position 
             of the susceptibility are shown by filled circles.
}
\label{fig:dcf}
\vspace{-2mm}
\end{figure}

\setcounter{figure}{3}

\begin{figure}[p]
\begin{center}
    \leavevmode
    \epsfxsize=9.0cm 
    \epsfbox{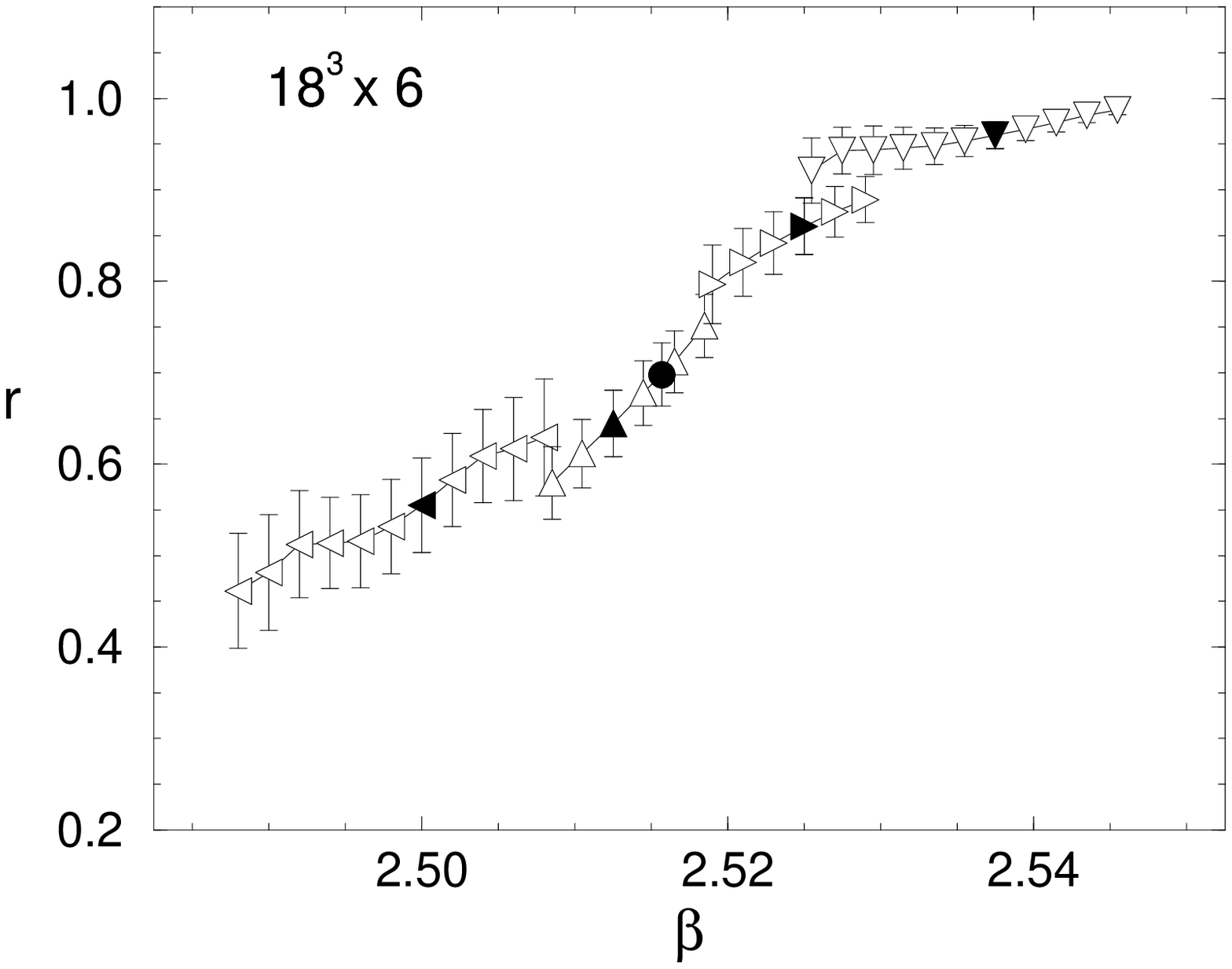}
\end{center}
    \caption{\it continued}
\end{figure}

\clearpage

\begin{figure}[t]
\vspace{-10mm}
\begin{center}
    \leavevmode
    \epsfxsize=14cm
    \epsfbox{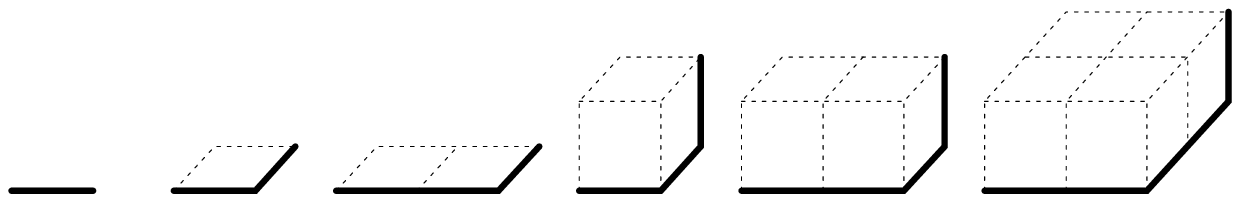}
\end{center}
\vspace{-8cm}
    \caption{Units of spatial paths of Wilson loops.}
    \label{fig:bus}
\vspace{-0.1cm}
\end{figure}

\clearpage

\begin{figure}[p]
\begin{center}
    \leavevmode
    \epsfxsize=9cm 
    \epsfbox{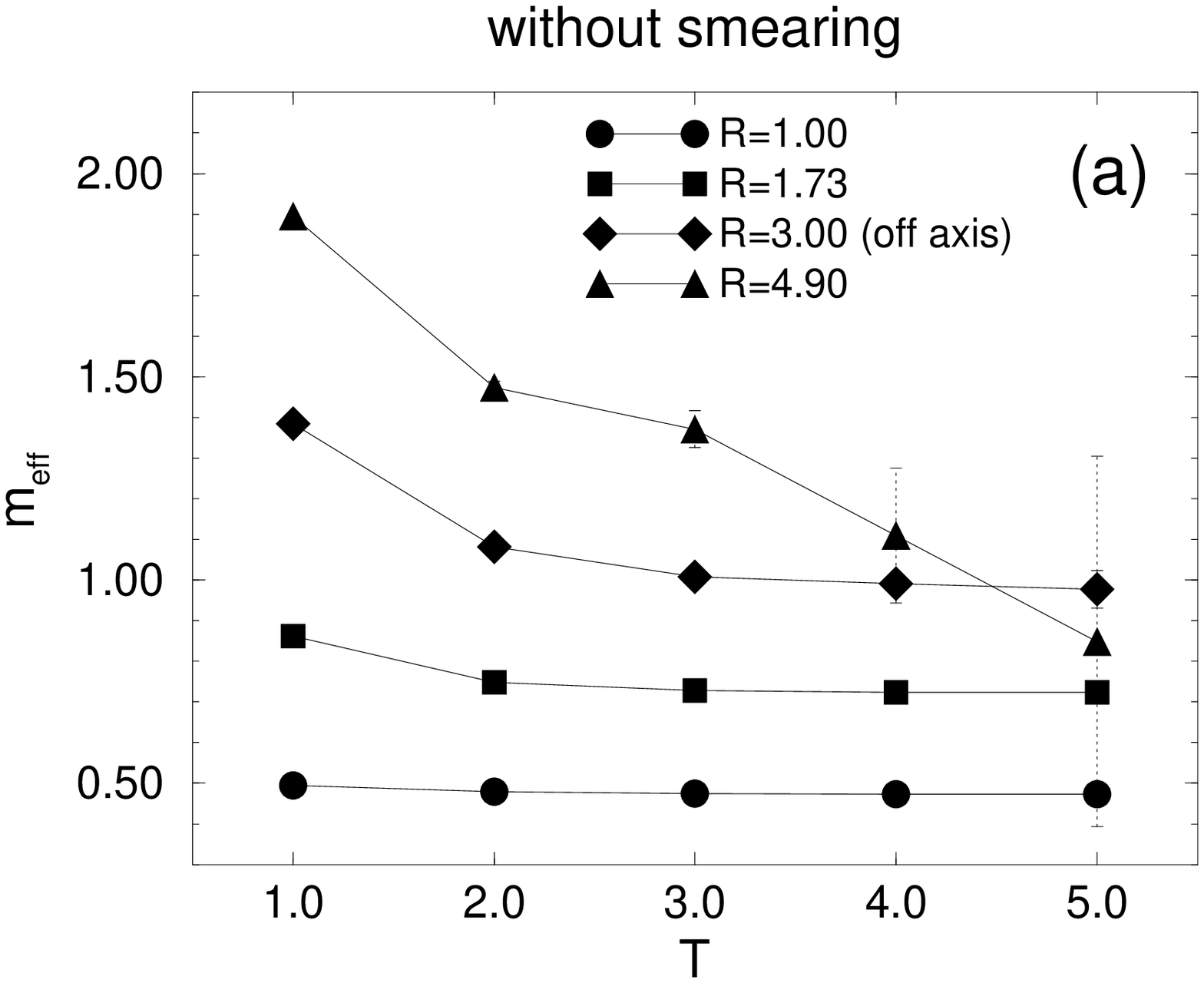}
\vspace{10mm}
    \epsfxsize=9cm 
    \epsfbox{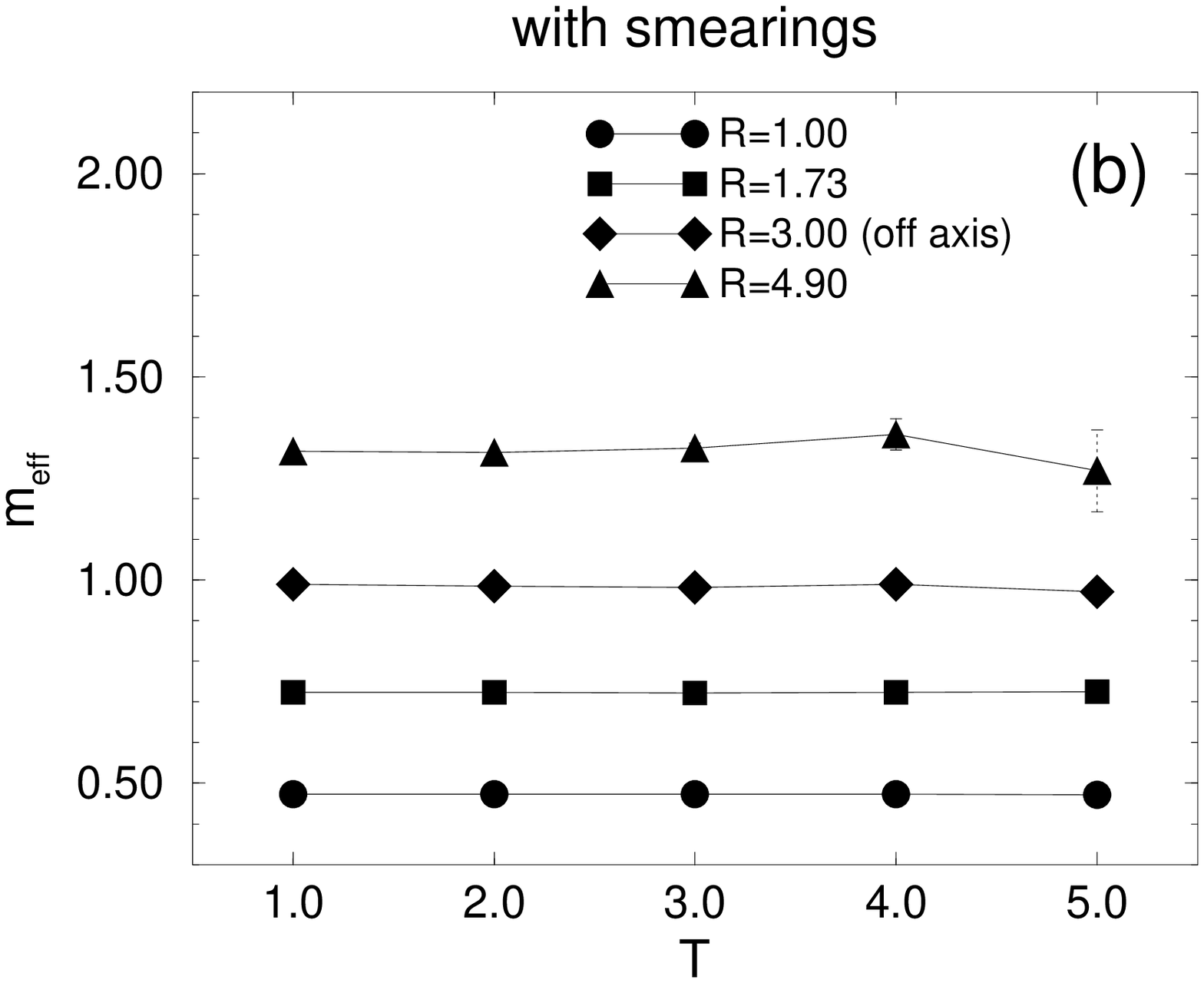}
\vspace{-0mm}
    \caption{$T$ dependence of $m_{\rm eff} = 
             \mbox{log}\left[W(R,T)/W(R,T+1)\right]$ 
             without (a) and with (b) smearing on the $12{^3}\times24$ 
             lattice.
             The results shown in (b) are obtained with the optimum 
             smearing step explained in the text.
             }
    \label{fig:em}
\end{center}
\vspace{0mm}
\end{figure}

\clearpage

\begin{figure}[t]
\begin{center}
\vspace{0mm}
    \leavevmode
    \epsfxsize=9cm 
    \epsfbox{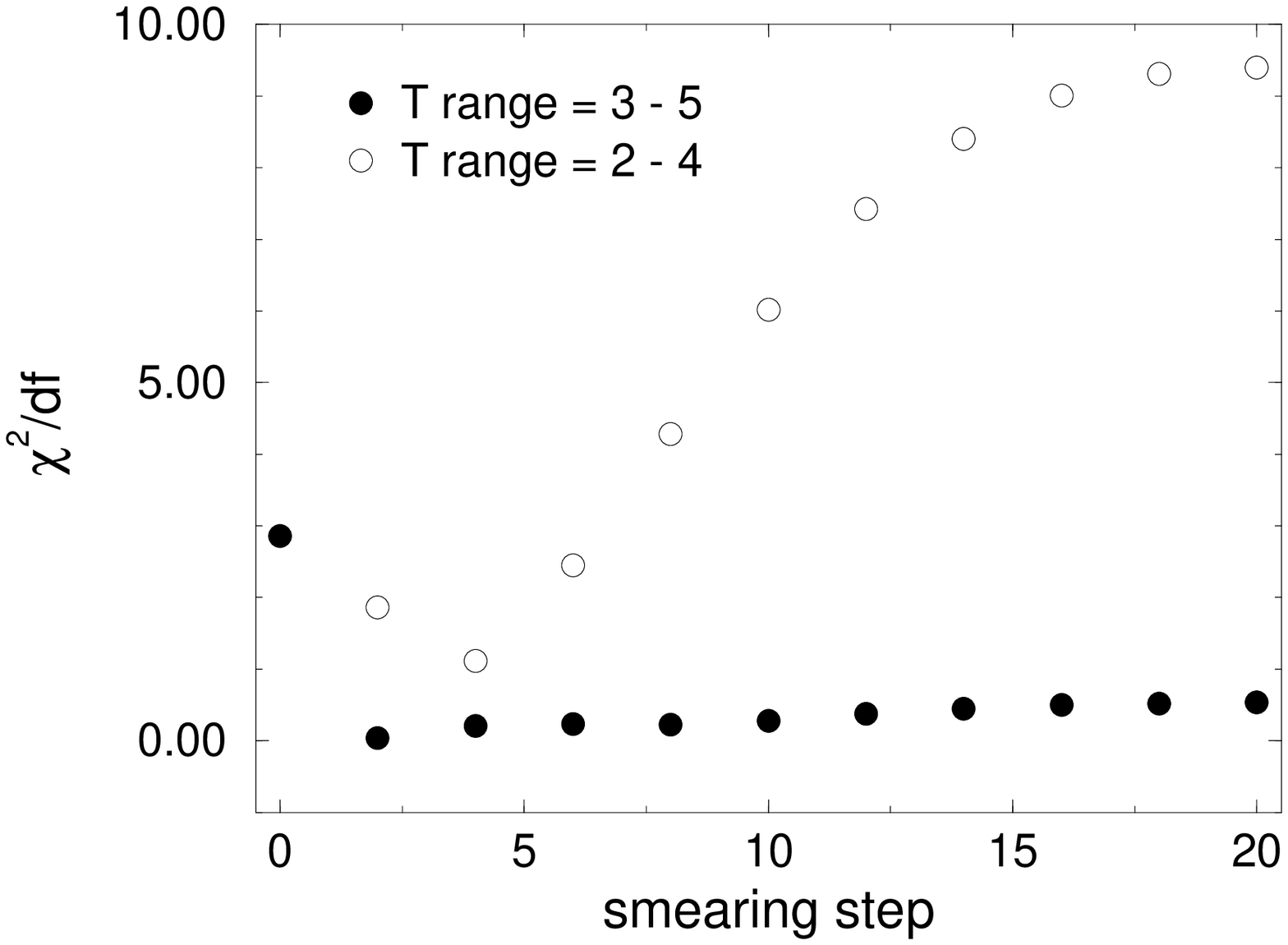}
\vspace{10mm}
    \epsfxsize=9cm 
    \epsfbox{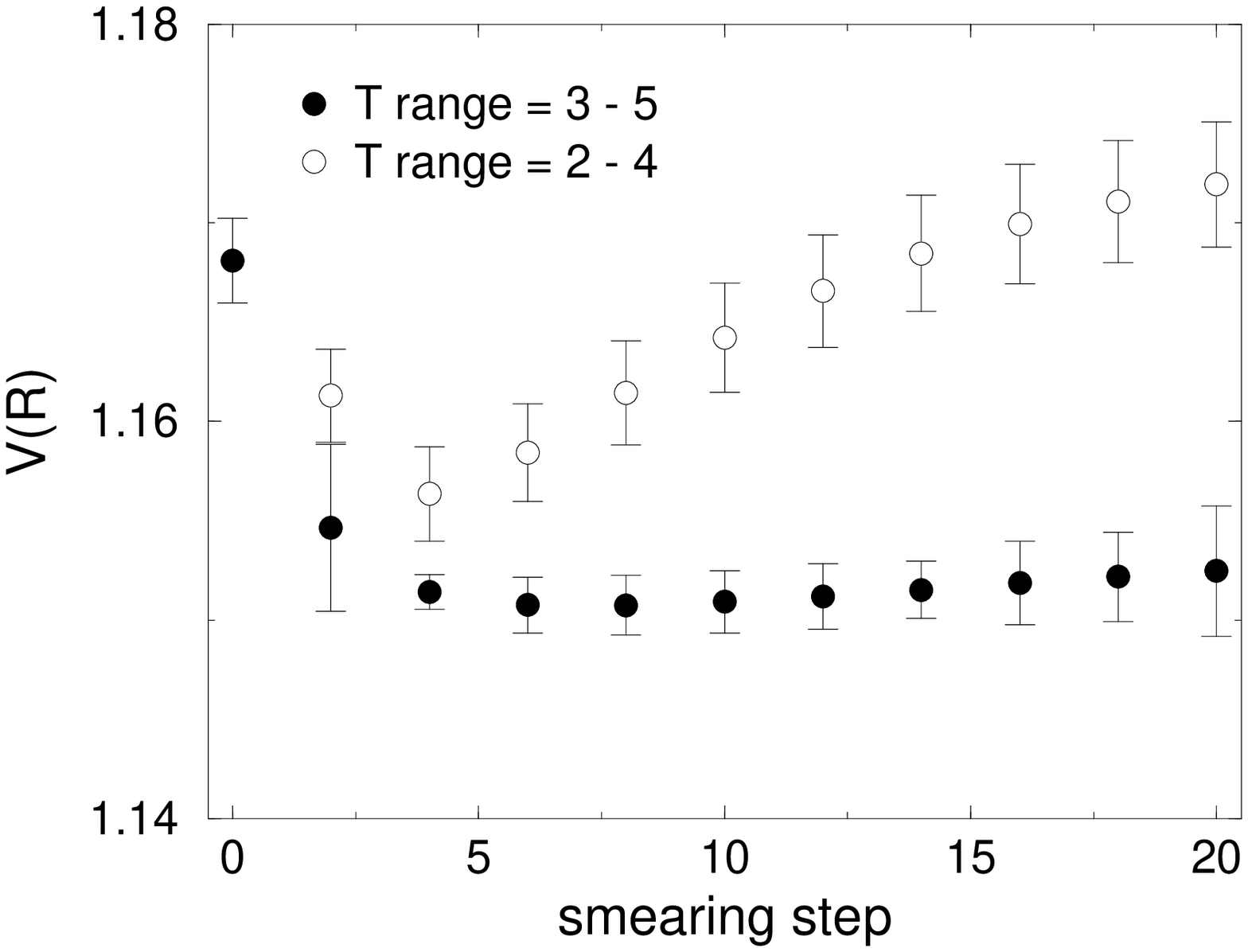}
\end{center}
\vspace{-0mm}
    \caption{Smearing step dependence of $\chi{^2}/df$ and $V(R)$ at
              $R=4.0$ on the $12^3 \times 24$ lattice.}
    \label{fig:cVstb}
\vspace{0cm}
\end{figure}

\clearpage

\begin{figure}[p]
\begin{center}
\vspace{-2mm}
\leavevmode
    \epsfxsize=10cm 
    \epsfbox{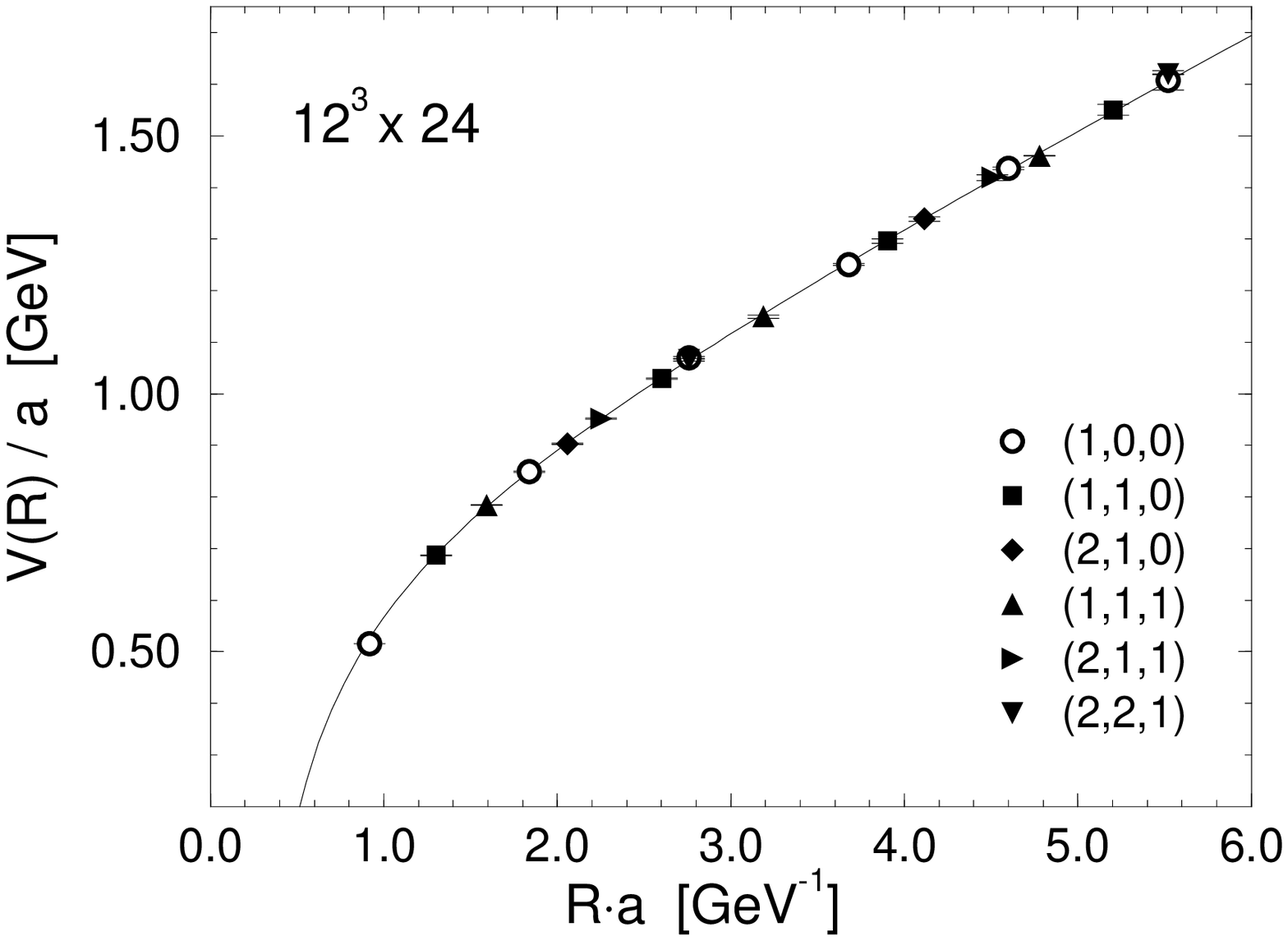}
\vspace{0mm}
    \epsfxsize=10cm 
    \epsfbox{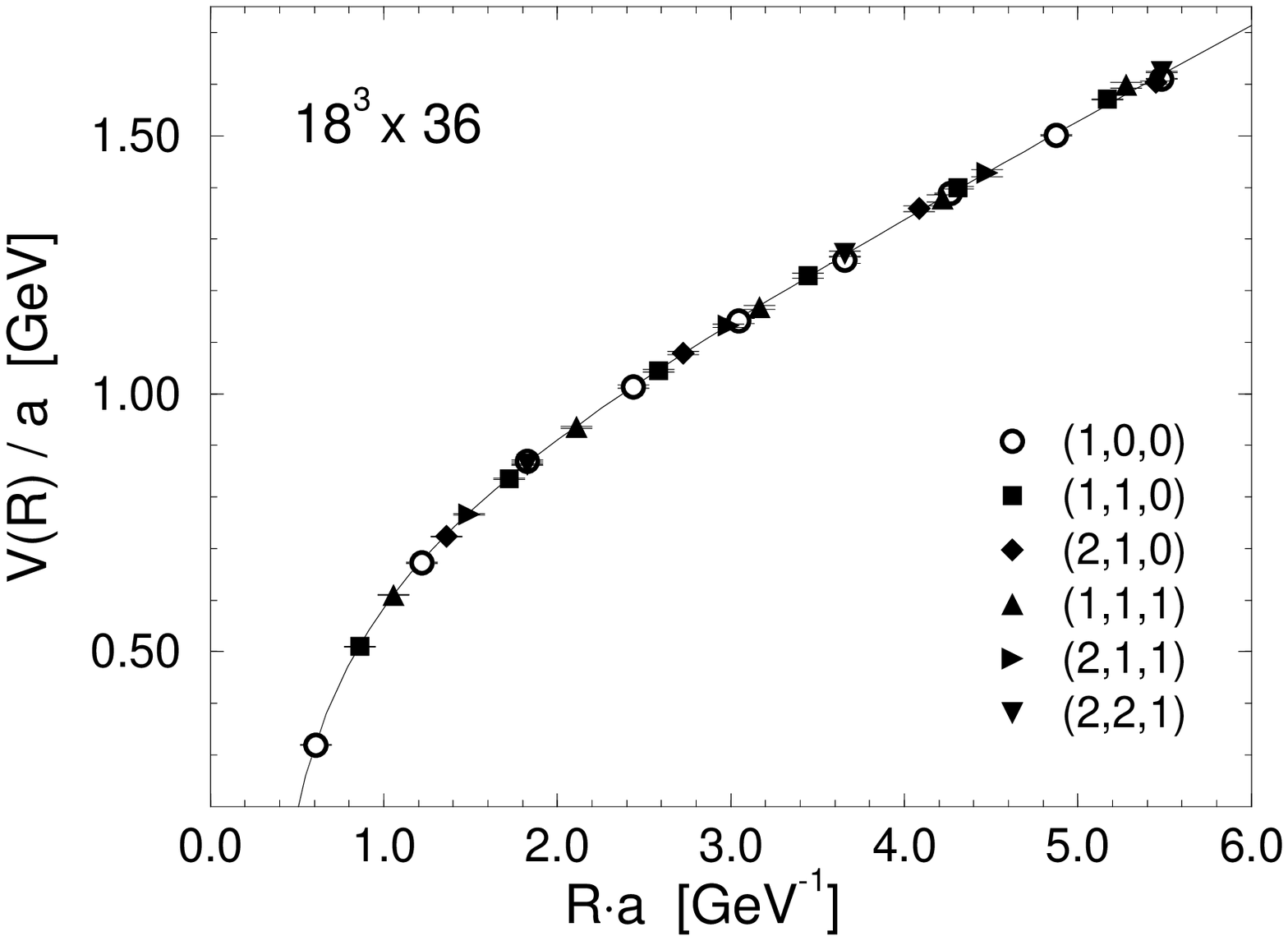}
\end{center}
\vspace{-0.5cm}
    \caption{Data for the potential and its fitting curves
             on the $12{^3}{\times}24$ 
             and $18{^3}{\times}36$ lattices. 
             The values of $a$ are determined from 
             $\sigma=(420\mbox{MeV})^2$.
	     The legends for the symbols represent
	     the units of spatial paths of Wilson loops 
             shown in Fig.~\protect\ref{fig:bus}. }
    \label{fig:VvsRds}
\vspace{-2mm}
\end{figure}

\clearpage

\begin{figure}[t]
\begin{center}
\vspace{5mm}
\leavevmode
    \epsfxsize=10cm 
    \epsfbox{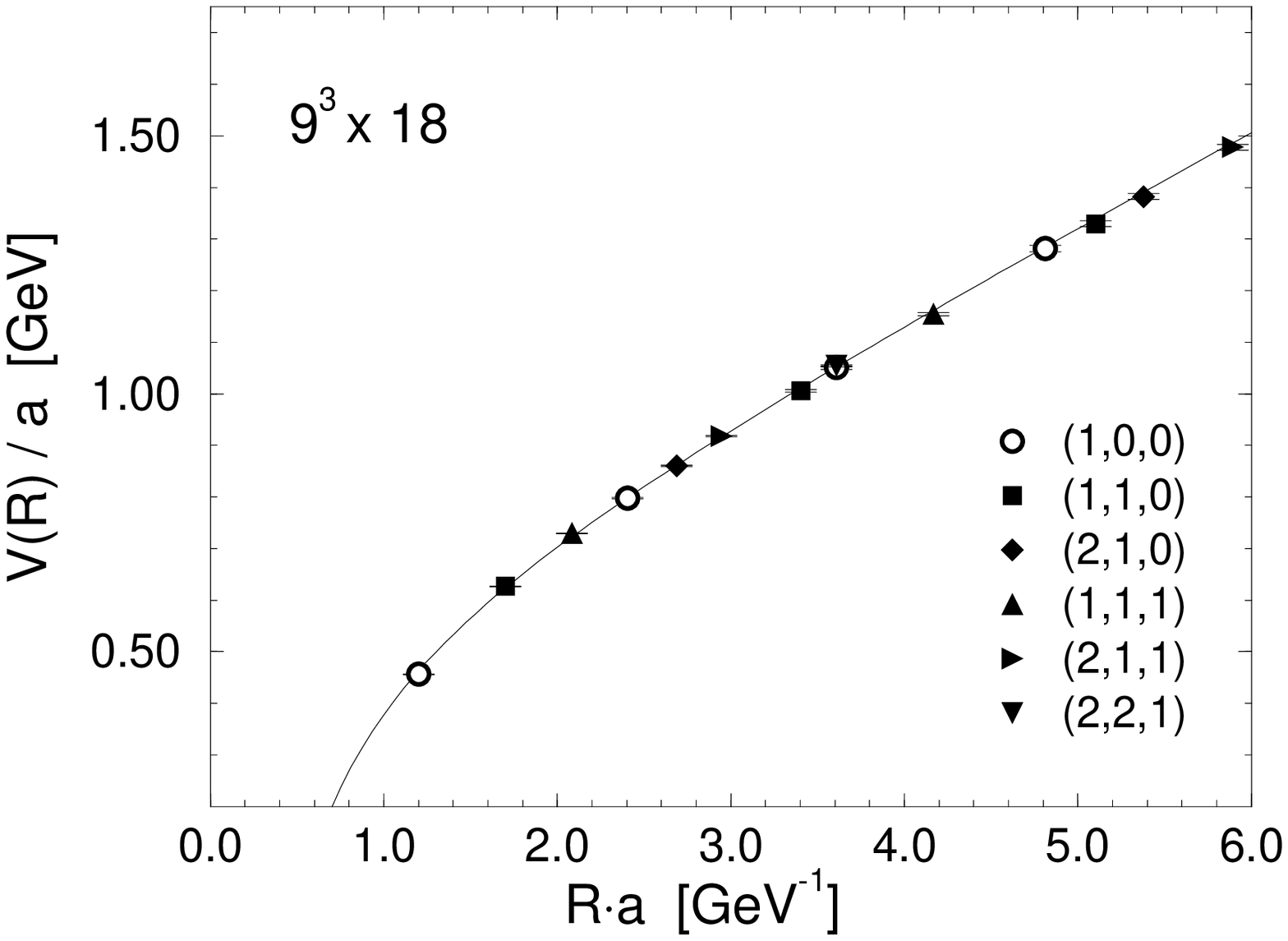}
\vspace{-0.2cm}
\end{center}
\vspace{-0.5cm}
    \caption{The same as Fig.~\protect\ref{fig:VvsRds} 
             obtained on the $9{^3}{\times}18$ lattice.}
    \label{fig:VvsRdsNt3}
\vspace{0cm}
\end{figure}

\clearpage

\begin{figure}[b]
\begin{center}
\leavevmode
    \epsfxsize=14cm 
    \epsfbox{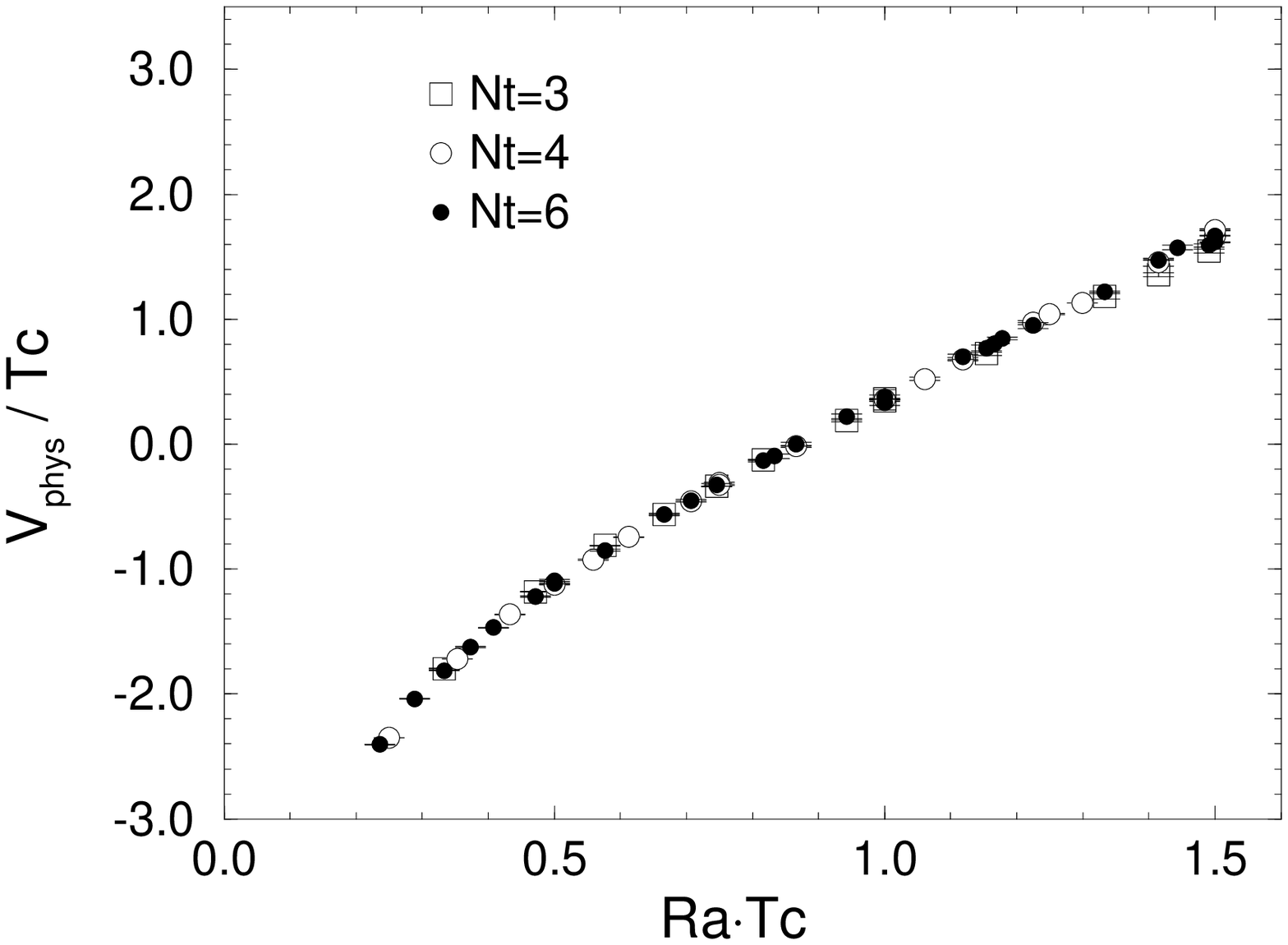}
\end{center}
\vspace{-0.7cm}
    \caption{$V_{\rm phys}/{T_c}$ vs.\ $Ra{\cdot}T_c$.
	     The constant term in the potential is fixed
	     so that the potentials have the same value
             at $Ra{\cdot}T_c=1.0$.}
\label{fig:VvsRTc}
\vspace{5mm}
\end{figure}

\clearpage

\begin{figure}[p]
\begin{center}
\vspace{-2mm}
\leavevmode
    \epsfxsize=10cm 
    \epsfbox{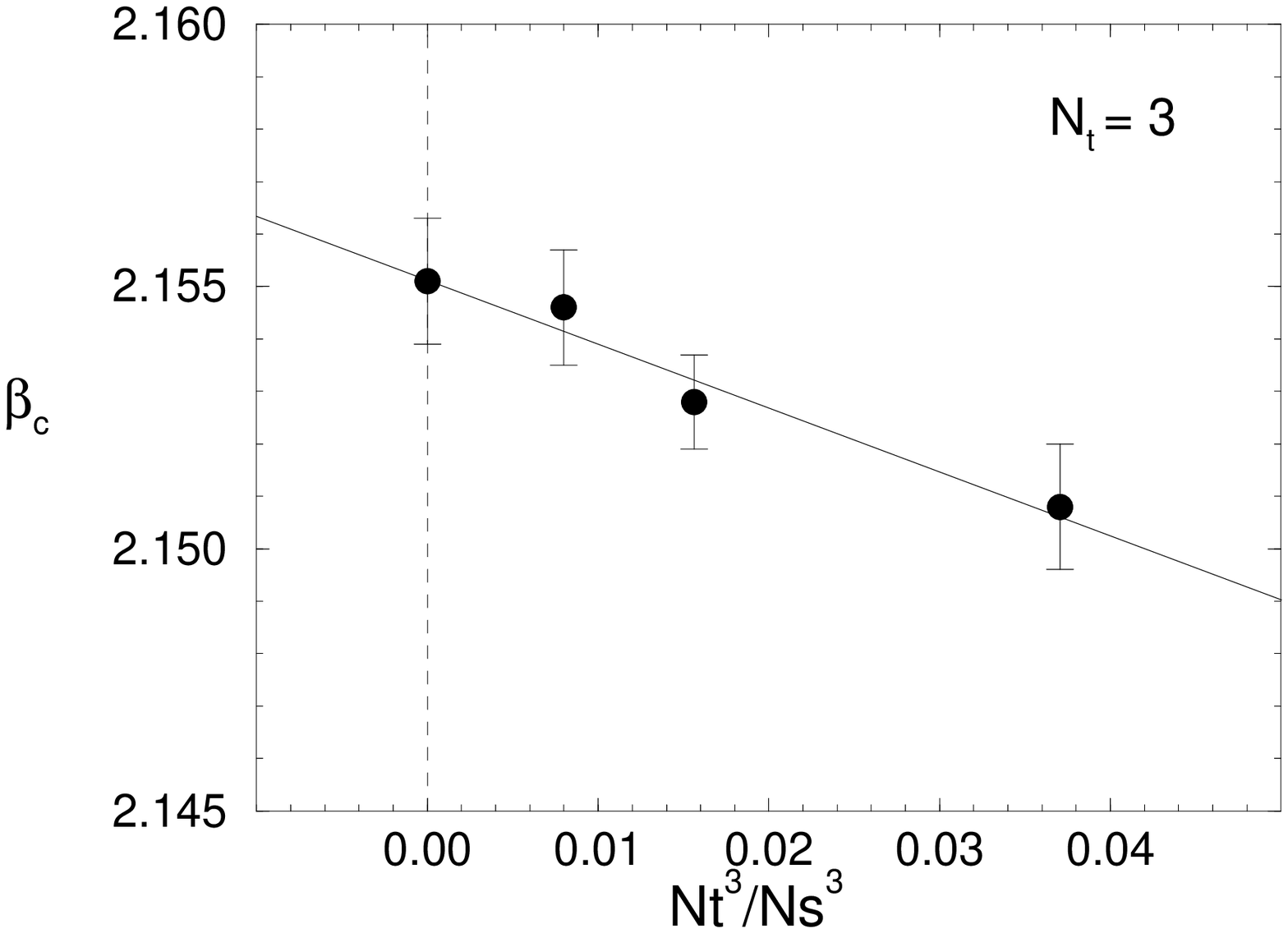}
\vspace{0mm}
    \epsfxsize=10cm 
    \epsfbox{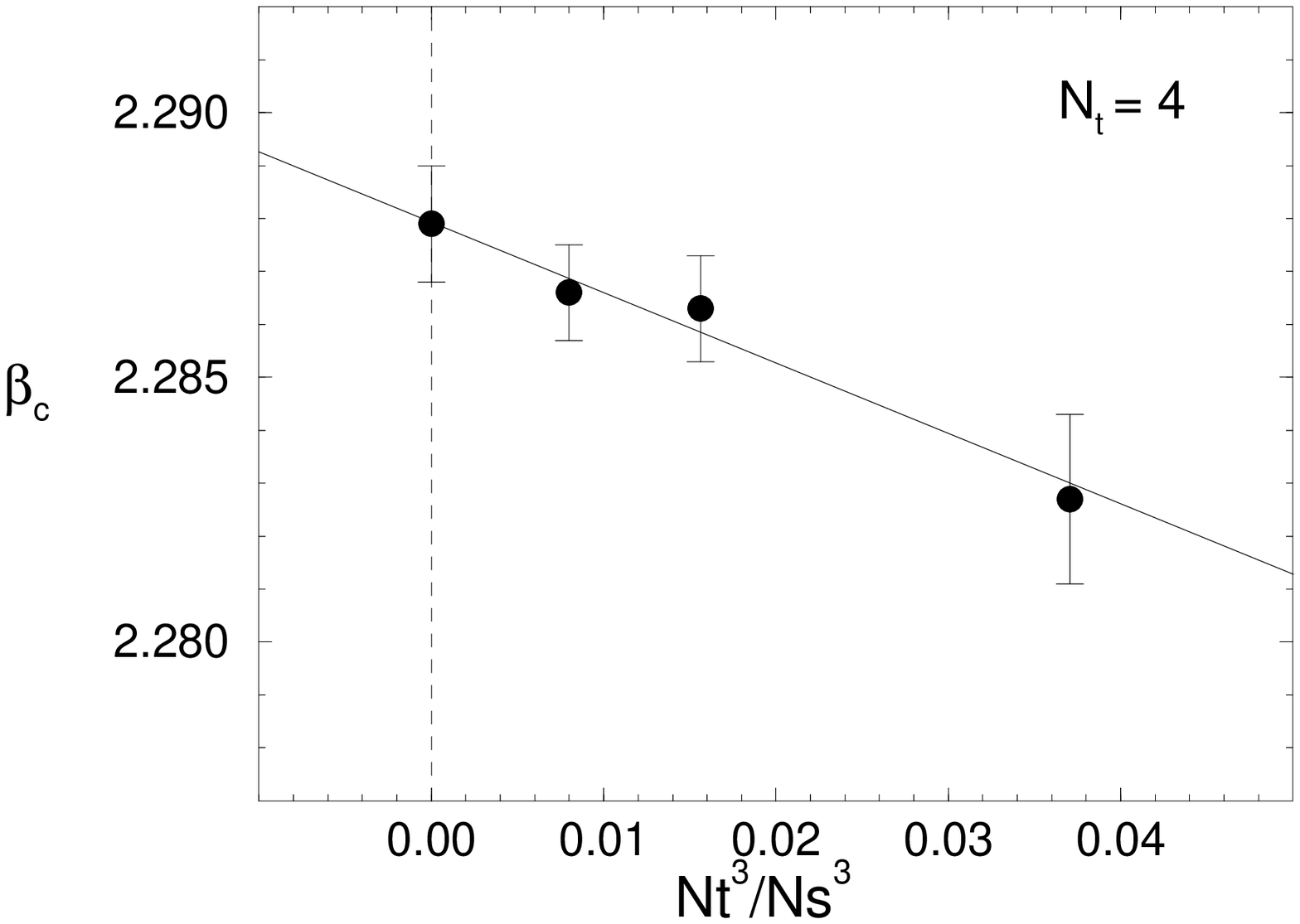}
\end{center}
\vspace{-0.5cm}
    \caption{Finite size scaling of $\beta_c$
             on $N_t=3$ and $4$ lattices.
             The solid lines are the results of a linear fit
             and the dashed lines indicate the location 
             of the infinite volume limit.}
    \label{fig:FSSbc}
\vspace{-2mm}
\end{figure}

\clearpage

\begin{figure}[t]
\begin{center}
\leavevmode
    \epsfxsize=14cm 
    \epsfbox{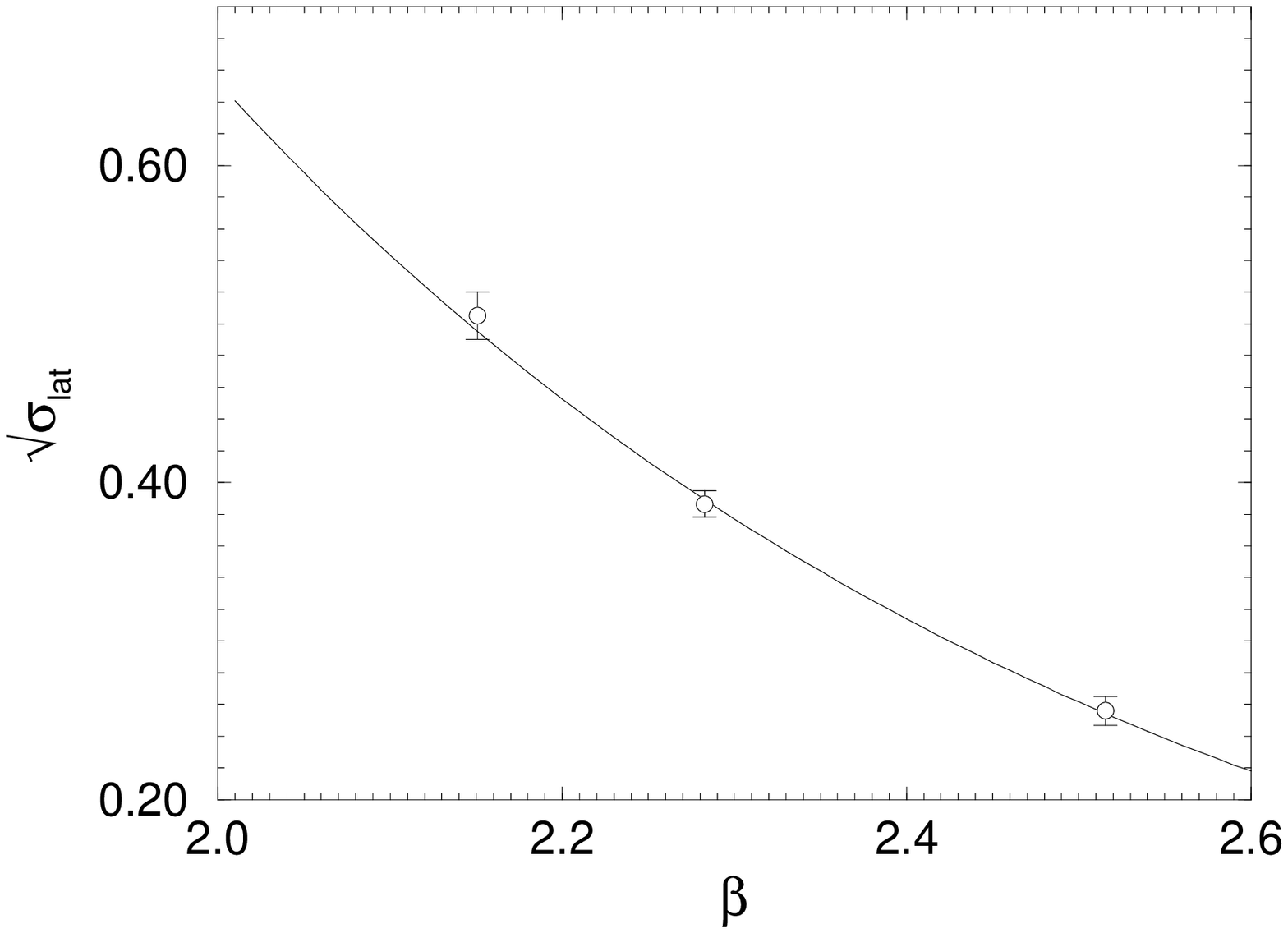}
\end{center}
\vspace{-0.7cm}
    \caption{Square root of the string tension
             ${\protect\sqrt\sigma_{\rm lat}}$
             at $\beta_c$ on the finite volume lattices 
	     simulated together with its fitting curve.
             }
\label{fig:Sigma}
\vspace{5mm}
\end{figure}

\clearpage

\begin{figure}[t]
\begin{center}
    \leavevmode
    \epsfxsize=14cm 
    \epsfbox{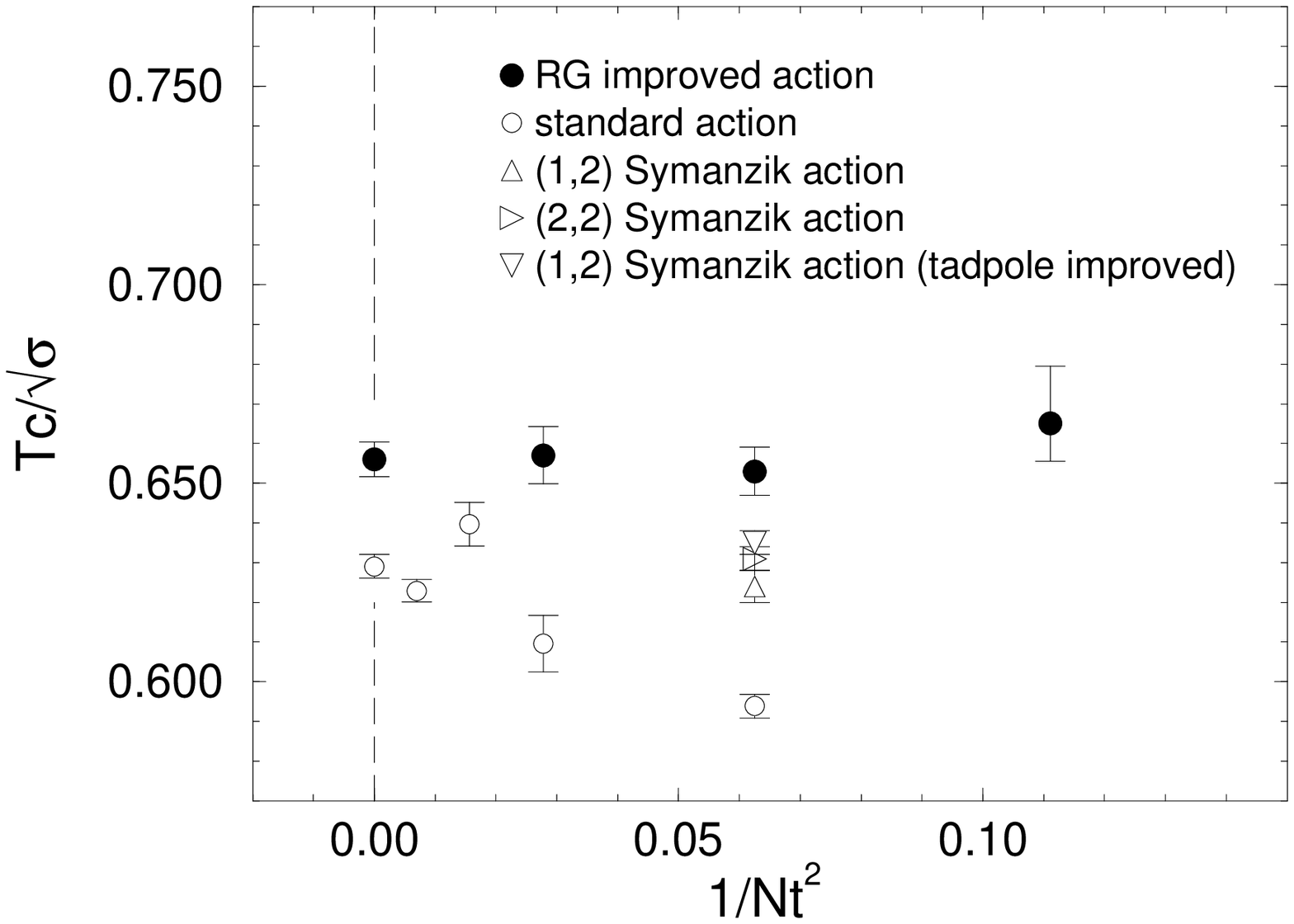}
\end{center}
\vspace{-7mm}
    \caption{$T_c/{\protect\sqrt\sigma}$ vs.\ $1/{{N_t}^2}$.
             All symbols represent the values 
             in the infinite volume limit.
The errors for our data in the cases of $N_t=4$ and 6 are statistical, 
while those
for $N_t=3$ include the systematic error due to uncertainty of the
fitting range.}
    \label{fig:Tcs}
\vspace{0mm}
\end{figure}

\end{document}